\documentclass[aps,twocolumn,twoide,a4paper,10pt,amsmath,amssymb,superscriptaddress]{revtex4}

\usepackage{mathrsfs}
\usepackage{graphicx}
\usepackage{amsmath}
\usepackage{amssymb}
\usepackage{amsfonts}
\usepackage{color}
\usepackage{amsmath}
\usepackage{CJK}
\usepackage{leftidx}

\def\bra#1{\left\langle{#1}\right|}
\def\ket#1{\left|{#1}\right\rangle}
\def\braket#1#2{\left\langle{{#1}}\mathrel{\left|{\vphantom{{#1}{#2}}}\right.\kern-\nulldelimiterspace}{{#2}}\right\rangle}

\begin{document}


\title{Many-body quantum lock-in amplifier}

\author{Min Zhuang}
\affiliation{Guangdong Provincial Key Laboratory of Quantum Metrology and Sensing $\&$ School of Physics and Astronomy, Sun Yat-Sen University (Zhuhai Campus), Zhuhai 519082, China}
\affiliation{State Key Laboratory of Optoelectronic Materials and Technologies, Sun Yat-Sen University (Guangzhou Campus), Guangzhou 510275, China}

\author{Jiahao Huang}
\altaffiliation{Email: hjiahao@mail2.sysu.edu.cn, eqjiahao@gmail.com}
\affiliation{Guangdong Provincial Key Laboratory of Quantum Metrology and Sensing $\&$ School of Physics and Astronomy, Sun Yat-Sen University (Zhuhai Campus), Zhuhai 519082, China}

\author{Chaohong Lee}
\altaffiliation{Email: lichaoh2@mail.sysu.edu.cn, chleecn@gmail.com}
\affiliation{Guangdong Provincial Key Laboratory of Quantum Metrology and Sensing $\&$ School of Physics and Astronomy, Sun Yat-Sen University (Zhuhai Campus), Zhuhai 519082, China}
\affiliation{State Key Laboratory of Optoelectronic Materials and Technologies, Sun Yat-Sen University (Guangzhou Campus), Guangzhou 510275, China}

\begin{abstract}
  Achieving high-precision detection of time-dependent signals in noisy environment is a ubiquitous issue in physics and a critical task in metrology.
  Lock-in amplifiers are detectors that can extract alternating signals with a known carrier frequency from an extremely noisy environment.
  Here, we present a protocol for achieving an entanglement-enhanced lock-in amplifier via empoying many-body quantum interferometry and periodic multiple pulses.
  Generally, quantum interferometry includes three stages: initialization, interrogation, and readout.
  The many-body quantum lock-in amplifier can be achieved via adding suitable periodic multiple-$\pi$-pulse sequence during the interrogation.
  Our analytical results show that, by selecting suitable input states and readout operations, the frequency and amplitude of an unknown alternating field can be simultaneously extracted via population measurements.
  In particular, if we input spin cat states and apply interaction-based readout operations, the measurement precisions for frequency and amplitude can both approach the Heisenberg limit.
  Moreover, our many-body quantum amplifier is robust against extreme stochastic noises.
  Our study may point out a new direction for measuring time-dependent signals with many-body quantum systems, and provides a feasible way for achieving Heisenberg-limited detection of alternating signals.
\end{abstract}
\date{\today}

\maketitle

\section{Introduction\label{Sec1}}

Detecting time-dependent alternating signals with high precision is an important problem in fundamental physics and a critical task in metrology~\cite{CWHelstrom1976,SLBraunstein1994,VGiovannetti2006,BMEscher2011,RDemkowiczDobrz2012,CLDegen2017,Vengalattore2007,Lee2006,Lee2009}.
Generally, the target signal is submerged in a noisy environment.
To obtain a high signal-to-noise ratio of the target signal, one has to choose an optimal tradeoff between decreasing the effect of noise and enhancing the response to the target signal.
Usually, lock-in amplifier has been used as a high-efficient classical detector to extract a signal from noisy environments with high signal-to-noise ratio.
The lock-in amplifier selects the signal with a specific frequency and phase from a mixture of mostly unwanted frequencies.

For quantum probes, dynamical decoupling (DD) methods have been extensively used to improve the signal-to-noise ratio for alternating signals~\cite{GdeLange2010,WJKuo2011,LJiang2011,PZanardi2008,PCMaurer2012,HStrobel2014,MSkotiniotis2015,Hosten2016,JGBohnet2016,ILovchinsky2016,SChoi2017,Biercuk2009,Hirose2012,JMBossl2017,JRMaze2008,GBalasubramanian2008,GdeLange2011,SKotler2011,RShaniv2017,SSchmitt2017}.
%
%
%
%
In particular, a quantum analogue to the classical lock-in amplifier with a single trapped Sr$^+$ ion has been demonstrated~\cite{SKotler2011}, in which several non-commuting dynamical manipulations to the spin states are performed to decouple the quantum probe from noise while enhancing its sensitivity to alternating signals.
The quantum lock-in techniques are readily available for other quantum probes, which can be useful for precision measurements.
However, to our best of knowledge, all existed studies of quantum lock-in amplifiers concentrate on single-particle systems.
Can one achieve a quantum lock-in amplifier via a many-body quantum system?

\begin{figure*}[!htp]
 \includegraphics[width=1.75\columnwidth]{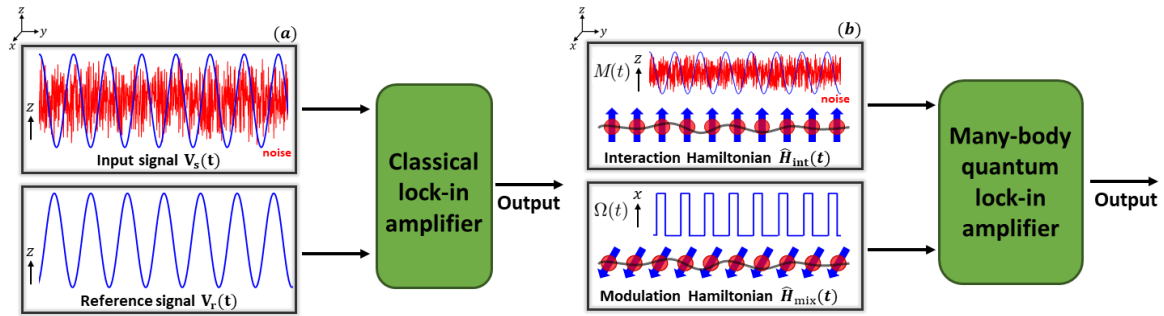}
  \caption{\label{Fig1}(color online).
  The schematic of classical lock-in amplifier and many-body quantum lock-in amplifier.
  (a) The classical lock-in amplifier.
  $V_{s}(t)$ is the target signal submerged in an extremely noisy environment.
  $V_{r}(t)$ is a known reference signal.
  Inputting the two signals through the classical lock-in amplifier, the signal $V_{s}(t)$ will be obtained.
  (b) The many-body quantum lock-in amplifier.
  The coupling between the probe and the signal is described by ${\hat{H}}_\textrm{int}=\emph{M}(t) \hat{J}_{z}$ with $\emph{M}(t)=\emph{S}(t)+\emph{N}_{o}(t)$, the target signal $\emph{S}(t)$ and the stochastic noise $\emph{N}_{o}(t)$.
  The oscillating modulation term ${\hat{H}}_\textrm{mix}$, which is analogue to the reference signal $V_{r}(t)$, does not commute with ${\hat{H}}_\textrm{int}$.
  Thus the probe evolution, obeying the Hamiltonian ${\hat{H}}={\hat{H}}_\textrm{int}+{\hat{H}}_\textrm{mix}$, can be used to realize a quantum lock-in amplifier for extracting the signal $\emph{S}(t)$.
  }
\end{figure*}

It is well known that, many-body quantum entanglement is a useful resource for improving measurement precision~\cite{VGiovannetti2004,VGiovannetti2011,JHuang2014,JGBohnet2016}.
For $N$ individual particles, the measurement precision only scales as the standard quantum limit (SQL) (i.e. $\propto 1/\sqrt{N}$).
The SQL can be surpassed by using entangled particles.
Remarkably, entangled non-Gaussian states (ENGSs), such as spin cat states or even Greenberger-Horne-Zeilinger (GHZ) state, set a benchmark for beating the SQL in metrology.
The measurement precision can be improved to the Heisenberg limit (i.e. $\propto 1/N$)~\cite{JJBollinger1996,TMonz2011,JHuang2015,SDHuver2008,LPezze2009}.
With state-of-art techniques, the spin cat states can be generated in ultracold atomic gases~\cite{HStrobel2014,BLucke2011,Gabbrielli2015,Bookjans2011,BLucke2011,JLHelm2018,HXin2016,JHuang2018032116}, trapped ions~\cite{Monz2011}, and photonic systems~\cite{JPan2012}.
Can one employ many-body quantum entanglement to improve the measurement precision of a quantum lock-in amplifier?
In further, whether the entanglement-enhanced lock-in amplifier is robust against environment noises?

In this article, we present a general protocol for achieving a many-body quantum lock-in amplifier via quantum interferometry and periodic multiple pulses.
In many-body quantum systems, the interferometry can be divided into three steps: state preparation, signal interrogation and readout.
To detect an unknown alternating signal, we apply a train of $\pi$-pulses with equidistant spacing in the signal interrogation stage.
The modulation (i.e. the applied $\pi$-pulse sequence) is non-commuting with the signal acquisition, which is the key for achieving quantum lock-in amplifier.
The pulse spacing $\tau=\pi/\omega_e$ is adjusted in order to probe the oscillating frequency $\omega$ of the target signal.
At resonance $\omega_e=\omega$, when the signal and the modulation pulses are locked in, the $\pi$ pulses are applied at every peak and valley of the signal and thus the accumulated phase is very close to zero.
While for $\omega_e\neq\omega$, the accumulated phase oscillates dramatically and is sensitively dependent on the detuning $\omega-\omega_e$.
As this accumulated phase determines the half-population, one may infer the resonant point $\omega_e=\omega$ via measuring the half-population versus the detuning.
Meanwhile, the signal amplitude can also be extracted.

Moreover, we analytically study the measurement precisions for frequency and amplitude of an AC magnetic field via many-body quantum systems.
For non-entangled states, the measurement precisions only attain the SQL.
Through utilizing entangled states, the measurement precisions can be improved to the Heisenberg limit.
In particular, by employing spin cat state and suitable interaction-based readout operations~\cite{EDavis2016,TMacri2016,FFrowis2016,Szigeti2017,Nolan2016,JHuang2018012129,Mirkhalaf2018,Anders2018,Burd2019,Linnemann2016,OHosten2016}, the measurement precisions for frequency and amplitude can simultaneously achieve the Heisenberg-limited scaling.
Besides, our many-body quantum amplifiers are robust against stochastic noises, even with highly entangled spin cat states.

This paper is organized as follows.
In Sec.~\ref{Sec2}, we introduce the general protocol for our many-body quantum lock-in amplifier.
In Sec.~\ref{Sec3}, we show how to achieve a many-body quantum lock-in amplifier with individual particles.
In Sec.~\ref{Sec4}, we demonstrate how to achieve the Heisenberg-limited quantum lock-in amplifier via spin cat states.
In Sec.~\ref{Sec5}, we discuss the robustness of our many-body quantum lock-in amplifier against stochastic noise.
Finally, we give a brief summary and discussion in Sec.~\ref{Sec6}.

\section{General protocol \label{Sec2}}

Classical lock-in amplifiers can extract a signal with a known carrier frequency from a noisy environment, see Fig.~\ref{Fig1}(a).
In essence, a lock-in amplifier receives a target signal, multiplies it by a reference signal, and integrates over a specified time.
After the integration, there is almost no contribution from any signal that is not at the same frequency as the reference signal.
%
%
In analogy, a quantum lock-in amplifier with many-body quantum systems can be realized by using quantum control techniques, see Fig.~\ref{Fig1}(b).
For a quantum lock-in amplifier, non-commutativity of different operations plays an important role.
%

To show the general protocol, we consider an ensemble of $N$ identical two-state bosonic particles.
The two states can be selected as any desired two levels, and hereafter we label them as spins $\ket{\uparrow}$ and $\ket{\downarrow}$, respectively.
The system states can be well characterized by the collective spin operators:
$\hat{J}_{x}=\frac{1}{2}(\hat{a}^{\dag}\hat{b}+\hat{a}\hat{b}^{\dag}),\hat{J}_{y}=\frac{1}{2i}(\hat{a}^{\dag}\hat{b}-\hat{a}\hat{b}^{\dag}),
\hat{J}_{z}=\frac{1}{2}(\hat{a}^{\dag}\hat{a}-\hat{b}^{\dag}\hat{b})$,
where $\hat{a}$ and $\hat{b}$ denote annihilation operators for spins $\ket{\uparrow}$ and $\ket{\downarrow}$, respectively.
The system states can be represented in terms of the Dike basis $\{|J,m\rangle\}$, where $\hat J_z |J,m\rangle = m|J,m\rangle$ with $J=\frac{N}{2}$ and $m = -J,-J + 1, ..., J-1, J$.
The interaction between the system and the external AC field is described by the Hamiltonian, ${\hat{H}}_\textrm{int}(t)=M(t) \hat{J}_{z}$, where the external field $M(t)=S(t)+N_{o}(t)$ consists of the target signal $S(t)=S_{0}\sin(\omega t+\beta)$ and the stochastic noise $\emph{N}_{o}(t)$.
The target signal is modulated with time, $\omega$ corresponds to the oscillation frequency, $\beta$ is the initial phase of the signal, and $S_{0}$ stand for the strength of the signal.

To implement the quantum lock-in amplifier, we mix the system with an induced modulation signal that does not commute with ${\hat{H}}_\textrm{int}$.
Here, we consider the mixing term $\hat{H}_{\textrm{mix}}(t)=\Omega({t})\hat{J}_{x}$ and the interaction Hamiltonian becomes
\begin{equation}\label{H_int}
{\hat{H}}={\hat{H}}_\textrm{int}(t)+ \hat{H}_{\textrm{mix}}(t)= M(t) \hat{J}_{z} + \Omega({t})\hat{J}_{x}.
\end{equation}

Below we briefly illustrate how to achieve high signal-to-noise ratio.
The non-commutativity of the two modulation terms ${\hat{H}}_\textrm{int}$ and ${\hat{H}}_\textrm{mix}$ plays an important role for lock-in measurement.
In units of $\hbar=1$, the time-evolution of system state $\ket{\Psi(t)}$ obeys the Schr\"{o}dinger equation,
\begin{equation}\label{SE}
i\frac{\partial{\ket{\Psi(t)}}}{\partial{t}}=\left[M(t)\hat{J}_{z}+\Omega({t})\hat{J}_{x}\right]{\ket{\Psi(t)}}.
\end{equation}

In the interaction picture with respect to ${\hat{H}_{\textrm{mix}}}$ (see Supplementary Material), the time-evolution is described by,
\begin{eqnarray}\label{Eq:StatePartial}
i\frac{\partial{\ket{\Psi(t)}}_I}{\partial{t}}&=& \hat{H}_I(t){\ket{\Psi(t)}_I}\nonumber\\
&=& M(t)\left[\cos(\alpha)\hat{J}_{z}+\sin(\alpha)\hat{J}_{y}\right]{\ket{\Psi(t)}_I},
\end{eqnarray}
where $\ket{\Psi(t)}_I =e^{i\int^{t}_{0}{\hat{H}_{\textrm{mix}}(t')}\textrm{d}t'}{\ket{\Psi(t)}}$ and $\alpha=\int^{t}_{0} \Omega(t')dt'$.
At time $T$, the system state is
%
\begin{eqnarray}\label{Eq:StateEvolveI}
|\Psi(T)\rangle_I &=&\hat{\mathcal{T}} e^{-i\int_0^{T}M(t)\left[\cos(\alpha)\hat{J}_{z}+\sin(\alpha)\hat{J}_{y}\right]dt}|{{\Psi}}(0)\rangle_I\nonumber\\
&=& \hat{\mathcal{T}} e^{-i\left(\varphi_1\hat{J}_{z}+\varphi_2\hat{J}_{y}\right)}|{{\Psi}}(0)\rangle_I,
\end{eqnarray}
%
with the time-ordering operator $\hat{\mathcal{T}}$, the initial state $\ket{\Psi(0)}_I=|\Psi(0)\rangle$, and the two phase factors
\begin{eqnarray}\label{Eq:phi1}
\varphi_1&=& \int_0^{T}\left[S_{0}\sin(\omega t+\beta)+N_{o}(t)\right]\cos(\alpha)\textrm{d}t\nonumber\\
         &=& \int_0^{T}S(t)\cos(\alpha)\textrm{d}t + \int_0^{T} N_{o}(t)\cos(\alpha)\textrm{d}t,
\end{eqnarray}
and %
\begin{eqnarray}\label{Eq:phi2}
\varphi_2 &=&\int_0^{T}\left[S_{0}\sin(\omega t+\beta)+N_{o}(t)\right]\sin(\alpha)\textrm{d}t\nonumber\\
          &=&\int_0^{T}S(t)\sin(\alpha)\textrm{d}t + \int_0^{T} N_{o}(t)\sin(\alpha)\textrm{d}t.
\end{eqnarray}

If we apply a suitable modulation $\Omega(t)$ to make $\cos(\alpha)$ and $\sin(\alpha)$ periodic and synchronized with the signal $S(t)$, the phase accumulated owing to $S(t)$ can adds up coherently whereas the phase accumulated owing to stochastic noise $N_{o}(t)$ can be averaged away.
Especially, if the frequencies of the noise spectral components is far from the signal frequency $\omega$, the noise spectral components can be disappeared in the long-time integration.
In this way, the signal-to-noise ratio of the output can be significantly improved.
DD is one of the effective modulation methods for quantum control.
In the following, we will consider the mixing modulation $\Omega(t)$ by DD techniques and demonstrate how to use many-body entanglement to enhance the measurement precision.

\section{Quantum Many-body lock-in amplifier with non-entangled states}\label{Sec3}

\begin{figure*}[!htp]
 \includegraphics[width=1.75\columnwidth]{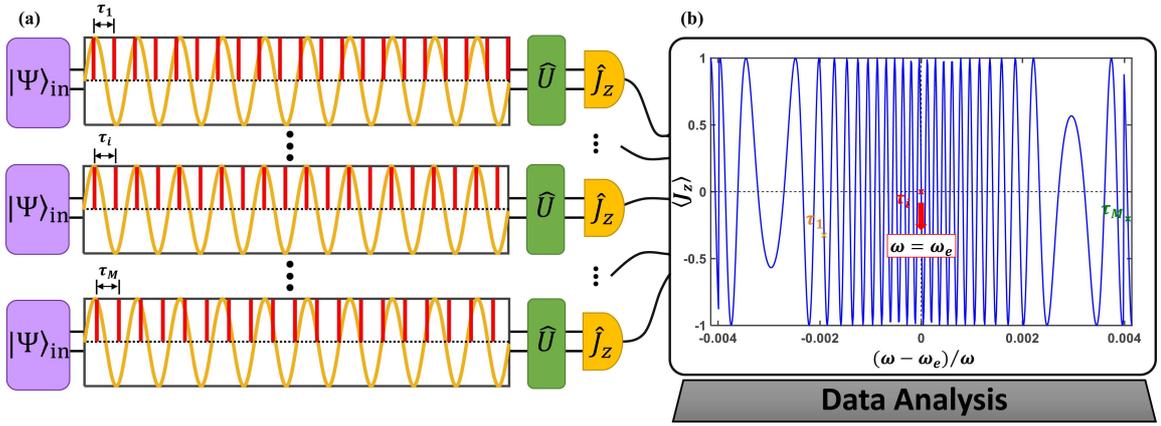}
  \caption{\label{Fig2}(color online).
  The general protocol of a many-body quantum lock-in amplifier for detecting an AC magnetic field.
  (a) The many-body quantum interferometry consists of three stages: (i) preparation, (ii) signal interrogation, and (iii) readout.
  In the initialization stage, an input state $|\Psi\rangle_\textrm{in}$ is prepared.
  Then, the input state undergoes an interrogation stage for signal accumulation.
  At this stage, the system state interacts with the magnetic field, and a train of $\pi$-pulses with specific equidistant spacing $\tau_i$ is applied.
  In the readout stage, a certain unitary operation $\hat{U}$ is applied for recombination, and the half-population difference is measured.
  Here, the pulse spacing $\tau_i$ ($i=1,2,...,M$) is adjusted in order to find the the frequency locking point $\omega=\omega_e$.
  (b) The half-population difference versus the detuning $\omega-\omega_e$ can tell the frequency locking point $\omega=\omega_e$.
  Here, for illustration, we choose a GHZ state with $N=2$, $T=4\pi$, $\omega=20\pi$ and $\gamma B=2\pi$.
  }
\end{figure*}

In this section, we will discuss how to use DD to achieve a many-body quantum lock-in amplifier for an AC magnetic field sensing.
Preparing the probe in non-entangled states, the measurement precisions can only attain the SQL.
While preparing the probe in entangled states, the measurement precisions can be improved to the Heisenberg limit, which will be discussed in Sec.~\ref{Sec4}.

Given the probe as an ensemble of two-mode bosonic particles, the coupling between the probe and the target signal [an AC magnetic field ${\textbf{B}}(t)=B\sin(\omega t+ \beta)\hat{\textbf{z}}$] is described by the Hamiltonian,
\begin{equation}\label{Eq:HamS}
\hat{H}_{B}(t)=\gamma {\textbf{B}}(t) \cdot \textbf{J}=\gamma B\sin(\omega t+ \beta)\hat{J}_{z}.
\end{equation}
Here, $B$ corresponds to the magnetic field amplitude, $\gamma$ is the gyromagnetic ratio, $\omega$ denotes the oscillation frequency, and $\beta$ is the initial phase.
For convenience, we assume $\beta=0$.
Our goal is to determine the frequency $\omega$ and the amplitude $B$ via many-body quantum lock-in amplifier.
Here, we neglect the stochastic noise $N_{o}(t)$ and its influences will be discussed in Sec.~\ref{Sec5}.

The key is to find out suitable modulation of $\Omega(t)$ for the mixing process.
In our scheme, the time-dependent modulation $\Omega(t)$ is designed as a sequence of $\pi$ pulses with equidistant spacing.
Generally, a single $\pi$ pulse can be expressed by a rectangular waveform, i.e., $\Omega_{\textrm{single}}(t)= \frac{\pi}{a}$ for $t_0\leq t \leq t_0+a$ with $a$ the width (duration) of the single $\pi$ pulse.
Ideally, the $\pi$ pulse is assumed sharp enough so that $a\rightarrow0$ and $\Omega_{\textrm{single}}(t)= \pi \delta(t-t_0)$, where $\delta(t)$ is the Dirac delta function.
For simplicity, we approximate $\Omega(t)$ as a sequence of equidistant sharp $\pi$ pulses so that it can be written as
\begin{eqnarray}\label{Eq:Omega}
\Omega(t)&=& \sum_{m=1}^{L} \Omega_{\textrm{single}}\left(t-(m+\frac{1+k}{2})\tau\right) \\\nonumber
         &=& \pi\sum_{m=1}^{L} \delta \left(t-(m+\frac{1+k}{2})\tau \right), \quad k\in[-1,1).
\end{eqnarray}
Here, $\tau$ is the spacing of the adjacent $\pi$ pulses, $M$ denotes the number of pulses.
We define the carrier frequency $\omega_e \equiv \frac{\pi}{\tau}$, which can be analogous to the carrier frequency for the classical lock-in amplifier.
$k$ determines the relative phase with respect to the target signal.
We consider $k=0$ in the following, which is assumed to be the lock-in condition.
When $\omega_e=\omega$, the $\pi$ pulses are applied at every peak and valley of the oscillating signal resulting in a tiny accumulated phase [i.e., Eq.~\eqref{Eq:phase1}], which can be used for frequency and phase locking.
Here, $k$ can be easily tuned by adjusting the start point of the initial pulse.
The other condition $k=-1$ will be discussed in Supplementary Material.

Through adjusting the carrier frequency $\omega_e$, we can extract the frequency $\omega$ of the AC magnetic field according to the population measurement, see Fig.~\ref{Fig2}.
For every fixed carrier frequency $\omega_e$, one can implement many-body quantum interferometry for measurement.

The many-body interferometry process can be divided into three stages: (i) probe preparation, (ii) signal interrogation, and (iii) readout, see Fig.~\ref{Fig2}.
In the initialization stage, an input state $|\Psi\rangle_\textrm{in}$ is prepared.
Then, the input state undergoes an interrogation stage for signal accumulation.
At this stage, the system state interacts with the AC magnetic field, and a train of $\pi$-pulses with an equidistant spacing are applied at the same time.
In the readout stage, a certain unitary operation $\hat{U}$ is applied for recombination, and the half-population difference is measured.

To adjust the carrier frequency $\omega_e$, one can choose $M$ different pulse spacing $\tau_i$, $i=1,2,...,M$.
The pattern of half-population difference measurement results versus $\omega-\omega_e$ can tell us the frequency locking point $\omega=\omega_e$, as shown in Fig.~\ref{Fig2}.
Now, we show how to find the frequency locking point.

We turn to discuss the signal interrogation stage.
When the evolution time $T$ is fixed, for the $i$-th pulse spacing, the number of the $\pi$ pulses is $L=[T/\tau_i+1/2]$, where $[\cdot]$ stands for the greatest integer.
Substituting Eq.~\eqref{Eq:Omega} into Eqs.~\eqref{Eq:StatePartial} and \eqref{Eq:StateEvolveI}, we obtain the Hamiltonian~\eqref{Eq:HamS} in the interaction picture
\begin{equation}\label{Eq:Ham1}
\hat{H}_\textrm{I}=h(t)\left[ \gamma B \sin(\omega t)\right]\hat{J}_{z},
\end{equation}
where $h(t)$ is the square wave function,
$$ h(t)=\left\{
\begin{array}{rcl}
1  &&{\omega_{e}t + \pi/2}\quad {\textrm{mod}\quad 2\pi\in[0,\pi)},\\
-1  &&{\omega_{e}t + \pi/2} \quad{ \textrm{mod}\quad 2\pi\in[\pi,2\pi).}
\end{array} \right. $$

In experiments, one can tune $\omega_e$ very close to $\omega$, that is, $|\omega-\omega_{e}|\ll\omega$.
Under the conditions of $|\omega-\omega_{e}|\ll\omega$ and $T\gg2\pi/\omega$, the Hamiltonian~\eqref{Eq:Ham1} can be approximated by an effective Hamiltonian,
\begin{equation}\label{Eq:Hameff1}
\hat{H}_{I}^{\textrm{eff}}(t)=\frac{2 \gamma B}{\pi}\sin\left[(\omega-\omega_{e}) t\right]\hat{J}_{z}.
\end{equation}
For an example, $T=2\pi$, the approximation is valid for $|\omega-\omega_{e}|/\omega < 0.1$ for $\omega\geq 10\pi$, see Supplementary Material.
The output state after the interrogation stage (of duration $T$) is given as, $\ket{\Psi}_{\textrm{out}}^{I} =e^{-i\int_{0}^{T}\hat{H}_{I}^{\textrm{eff}}(t)dt} \ket{\Psi}_{\textrm{in}}=e^{i\int^{T}_{0}{\hat{H}_{\textrm{mix}}(t)}\textrm{d}t}{\ket{\Psi}_{\textrm{out}}}=e^{i\alpha\hat{J}_x}{\ket{\Psi}_{\textrm{out}}}$, where $\ket{\Psi}_{\textrm{out}}$ is the output state in Schr\"{o}dinger picture.

Finally, in the readout stage, a suitable unitary operation $\hat{U}$ is performed for recombination.
The selection of unitary operation $\hat{U}$ depends on the input state and will have influences on the measurement precisions, which will be discussed in the next section.
The final state (in Schr\"{o}dinger picture) before half-population difference measurement can be written as
\begin{eqnarray}\label{Eq:final_state}
	 \ket{\Psi}_{\textrm{final}}&=&\hat{U}\ket{\Psi}_{\textrm{out}}=\hat{U}e^{-i\alpha\hat{J}_x}\ket{\Psi}_{\textrm{out}}^{I},\nonumber\\
&=&\hat{U}e^{-i\alpha\hat{J}_x}e^{-i\int_{0}^{T}\hat{H}_{I}^{\textrm{eff}}(t)dt} \ket{\Psi}_{\textrm{in}}.
\end{eqnarray}
Substituting Eq.~\eqref{Eq:Hameff1} into Eq.~\eqref{Eq:final_state}, we have $\ket{\Psi}_{\textrm{final}}=\hat{U}e^{-i\alpha\hat{J}_x}e^{-i\phi\hat{J}_z} \ket{\Psi}_{\textrm{in}}$.
Here, the accumulated phase reads
\begin{equation}\label{Eq:phase1}
	\phi=\frac{2\gamma B}{\pi}\frac{1-\cos\left[(\omega-\omega_{e}) T\right]}{\omega-\omega_{e}}.
\end{equation}
Thus, for the lock-in condition, when $\omega-\omega_e \rightarrow 0$, the accumulated phase $\phi \rightarrow \frac{\gamma B (\omega-\omega_e)T^2}{\pi}\approx 0$.

The expectation of the half-population difference measurement on final state is
\begin{eqnarray}\label{Eq:Expectation}
\langle\hat{J}_{z}\rangle_{\text{f}}&=&\!\bra{\Psi}_{\text{final}} \hat{J}_{z} \ket{\Psi}_{\text{final}} \nonumber\\
&=&\bra{\Psi}_{\textrm{out}}^{I} \hat{U}^{\dagger}_{I} \hat{J}_{z}^{I} \hat{U}_{I} \ket{\Psi}_{\textrm{out}}^{I} \nonumber\\
&=&\bra{\Psi}_{\textrm{final}}^{I} \hat{J}_{z}^{I}\ket{\Psi}_{\textrm{final}}^{I},
\end{eqnarray}
with $\hat{J}_{z}^{I}=e^{i\alpha\hat{J}_x} \hat{J}_z e^{-i\alpha\hat{J}_x}$ and $\hat{U}_{I}=e^{i\alpha\hat{J}_x} \hat{U} e^{-i\alpha\hat{J}_x}$.
At time $T$, we have $\alpha=L\pi$ and $\hat{U}_{I}=e^{iL\pi\hat{J}_x} \hat{U} e^{-iL\pi\hat{J}_x}$ .
When the pulse number $L$ is even, $\hat{J}_{x}^{I}=\hat{J}_x$, $\hat{J}_{y}^{I}=\hat{J}_y$, $\hat{J}_{z}^{I}=\hat{J}_z$; when $L$ is odd, $\hat{J}_{x}^{I}=\hat{J}_x$, $\hat{J}_{y}^{I}=-\hat{J}_y$, $\hat{J}_{z}^{I}=-\hat{J}_z$ (see Supplementary Material).
%

%
%

%
We first consider individual particles without any entanglement.
Suppose all the particles are prepared in the spin coherent state (SCS) $\ket{\Psi}_{\textrm{SCS}}=e^{-i\frac{\pi}{2}\hat J_y}\ket{N/2,-N/2}$.
This input state can be easily generated by applying a $\pi/2$ pulse on the state of all particles in spin-down $\ket{\downarrow}$.
In this situation, one can choose $\hat{U}_{I}=e^{-i\frac{\pi}{2}{\hat{J}_{y}}}$.
%
%
%
Then, the final state before the half-population difference measurement can be written as
\begin{eqnarray}\label{Evo_CSC}
|\Psi\rangle_{\text{final}}^{I}
=e^{-i\frac{\pi}{2}{\hat{J}_{y}}} e^{-i\phi\hat J_z}\ket{\Psi}_{\textrm{SCS}}.
\end{eqnarray}
In an explicit form, the final state becomes
\begin{eqnarray}\label{Evo_CSC}
|\Psi\rangle_{\text{final}}^{I}=\!\!\!\sum_{m=\!-J}^{J}\!\!\!\sqrt{C_{J}^{m}}\left[-\cos(\frac{\phi}{2})\right]^{\!J\!+\!m}\left[i\sin(\frac{\phi}{2})\right]^{\!J\!-\!m}\!\!\!|J,m\rangle. \nonumber \\
\end{eqnarray}
where $C_{J}^{m}={\frac{(2J)!}{(J+m)!(J-m)!}}$ is the binomial coefficient.
%
%
After some algebra, the expectations of half-population difference and the square of half-population difference on the final state can also be explicitly written as

\begin{eqnarray}\label{Jz_SCS}
\langle J_{z} \rangle_{\text{f}} &=& (-1)^L \frac{N}{2}\cos(\phi) \nonumber \\
&=&(-1)^L \frac{N}{2}\cos\left(\frac{2\gamma B}{\pi}\frac{1-\cos\left[(\omega-\omega_{e}) T\right]}{\omega-\omega_{e}}\right),
\end{eqnarray}
%
and
\begin{eqnarray}\label{Jz2_SCS}
\langle J_{z}^{2} \rangle_{\text{f}} =\frac{N}{4}+\frac{N(N-1)}{4}\cos^{2}(\phi).
\end{eqnarray}

\begin{figure}[!htp]
 \includegraphics[width=1\columnwidth]{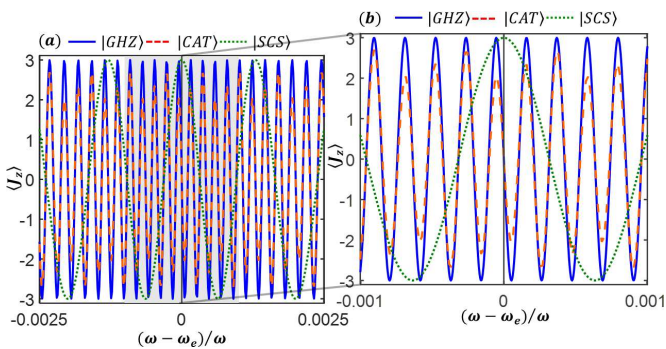}
  \caption{\label{Fig3}(color online).
  (a) The variation of half-population difference measurement versus detuning $\omega-\omega_{e}$.
  The green dot line are the results for a SCS with $\hat{U}_{I}=e^{-i\frac{\pi}{2}{\hat{J}_{y}}}$.
  The expectation of half-population difference is symmetric with respect to the zero detuning  $\omega-\omega_{e}=0$.
  The orange dashed line and blue solid line correspond to the results for the spin cat state  $\ket{\Psi(\theta=\pi/8)}_{\textrm{CAT}}$ and GHZ state $\ket{\Psi(\theta=0)}_{\textrm{CAT}}$, respectively, with $\hat{U}_{I}=e^{i\frac{\pi}{2}{\hat{J}_{x}}}e^{i\frac{\pi}{2}{\hat{J}_{z}^2}}e^{i\frac{\pi}{2}{\hat{J}_{x}}}$.
  For spin cat states, the expectation of half-population difference is anti-symmetric with respect to the zero detuning, i.e., $\langle\hat{J}_z\rangle =0$ for $\omega-\omega_e=0$.
  (b) The inset in the range of $(\omega-\omega_{e})/\omega \in [-0.001, 0.001]$.
  Here, $T=2\pi$, $\omega=20\pi$, $\gamma B=2\pi$, and $N=6$.
   }
\end{figure}

It is found that the information of the estimated two parameters $\omega$ and $B$ can be inferred from the oscillation of the half-population difference.
The form of Eq.~\eqref{Jz_SCS} is exactly symmetric with respect to detuning $\omega-\omega_{e}$, as shown in Fig.~\ref{Fig3} (green dotted line).

Thus, one can determine the zero detuning point $\omega-\omega_{e}=0$ [grey line in Fig.~\ref{Fig3}(a)] from the symmetry of the pattern for half-population difference.
Once the value of $\omega$ is determined, and one can further extract $B$ according to the analytic form of Eq.~\eqref{Jz_SCS} via a fitting procedure.
In this point, the values of $\omega$ and $B$ can be simultaneously obtained only via the half-population difference measurement in experiments.

%

To further characterize the measurement precision of $\omega$ and $B$, we use the error propagation formula~\cite{Helstrom1967,CWHelstrom1976,Paris2009,AdvPhysX2016}.
The measurement precisions for $\omega$ and $B$ are
\begin{equation}\label{Eq:omega uncertainty}
\Delta \omega=\frac{(\Delta{\hat{J}_{z}})_{\text{f}}}{|\partial{\langle\hat{J}_{z}\rangle_{\text{f}}}/ \partial{\omega_{e}}|},
\end{equation}
and
\begin{equation}\label{Eq:B uncertainty}
\Delta B=\frac{(\Delta{\hat{J}_{z}})_{\text{f}}}{|\partial{\langle\hat{J}_{z}\rangle_{\text{f}}}/ \partial{B}|},
\end{equation}
where
\begin{equation}\label{Eq:Deviation}
(\Delta{\hat{J}_{z}})_{\text{f}}=\sqrt{\langle\hat{J}_z^2\rangle_{\text{f}}-\langle\hat{J}_{z}\rangle_{\text{f}}^2}.
\end{equation}

Substituting Eq.~\eqref{Jz_SCS} and Eq.~\eqref{Jz2_SCS} into Eq.~\eqref{Eq:omega uncertainty} and Eq.~\eqref{Eq:B uncertainty}, one can analytically obtain $\Delta \omega$ and $\Delta B$.
They read as
\begin{equation}\label{SCS_Delta_omega0}
\Delta{\omega}\!=\!\frac{\pi}{2\gamma B\sqrt{N}}\!\frac{(\omega-\omega_{e})^2}{\left|G\right|}
\end{equation}
and
\begin{equation}\label{SCS_Delta_B0}
\Delta{B}=\frac{\pi}{2\gamma\sqrt{N}}\left|\frac{\omega-\omega_{e}}
{\cos\left[(\omega-\omega_{e}) T\right]-1}\right|,
\end{equation}
where
\begin{equation}
G=\cos\left[(\omega-\omega_{e}) T\right]-T(\omega-\omega_{e})\sin\left[(\omega-\omega_{e}) T\right]-1.
\end{equation}
From Eq.~\eqref{SCS_Delta_omega0} and Eq.~\eqref{SCS_Delta_B0}, the measurement precisions $\Delta{\omega}$ and $\Delta B$ for individual particles only exhibit the SQL scaling.
For a fixed $N$, the measurement precision $\Delta \omega$ is dependent on detuning $\omega-\omega_{e}$, the signal amplitude $B$ and evolution time $T$.
When $\omega-\omega_e \rightarrow 0$, $\Delta \omega \rightarrow \frac{\pi}{3\gamma B T^2\sqrt{N} }$ attains the optimal value.
However, the measurement precision $\Delta B$ is just dependent on detuning $\omega-\omega_{e}$ and $T$.
When $\omega-\omega_{e} \rightarrow 0$, $\Delta{B}\rightarrow \frac{\pi}{\gamma (\omega-\omega_{e})T^2\sqrt{N}}\rightarrow \infty$, which is diverged.
When $\cos[(\omega-\omega_e)T] + (\omega-\omega_e)T\sin[(\omega-\omega_e)T]=1$, $\Delta B$ attains its minimum.

We numerically find the optimal measurement precisions in the range of $\left|(\omega-\omega_{e})/\omega\right|\in[0,0.008]$ and calculate the corresponding scaling versus particle number for $\gamma B=2\pi$.
According to the fitting results (see Fig.~\ref{Fig4}), the log-log measurement precision for $\textrm{ln}\left({{\Delta \omega_{\textrm{min}}}/{\omega}}\right)\approx -0.5\textrm{ln}({N})-8.51$ and $\textrm{ln}\left({\gamma \Delta B_{\textrm{min}}}/{\omega}\right)\approx -0.5\textrm{ln}({N})-5.20$.
%
For the two parameters, the optimal measurement precision $\Delta \omega_{\textrm{min}}$ (purple circles) and $\Delta B_{\textrm{min}}$ (purple circles) both exhibit SQL scaling as expected.

\section{Quantum Many-body lock-in amplifier with spin cat states\label{Sec4}}
In this section, we will discuss how to realize the Heisenberg-limited measurement for $\omega$ and $B$ within this framework.
Entanglement is a useful quantum resource to improve the measurement precision.
Spin cat state, as a kind of non-Gaussian entangled state, is one of the promising candidates for Heisenberg-limited metrology~\cite{GFerrini2008,GFerrini2010,KPawlowski2013,JHuang2015,JHuang2018012129}.
%
%
%
%
In this section, we demonstrate how to use spin cat states to realize the Heisenberg-limited quantum many-body lock-in amplifier.

\begin{figure}[!htp]
 \includegraphics[width=1\columnwidth]{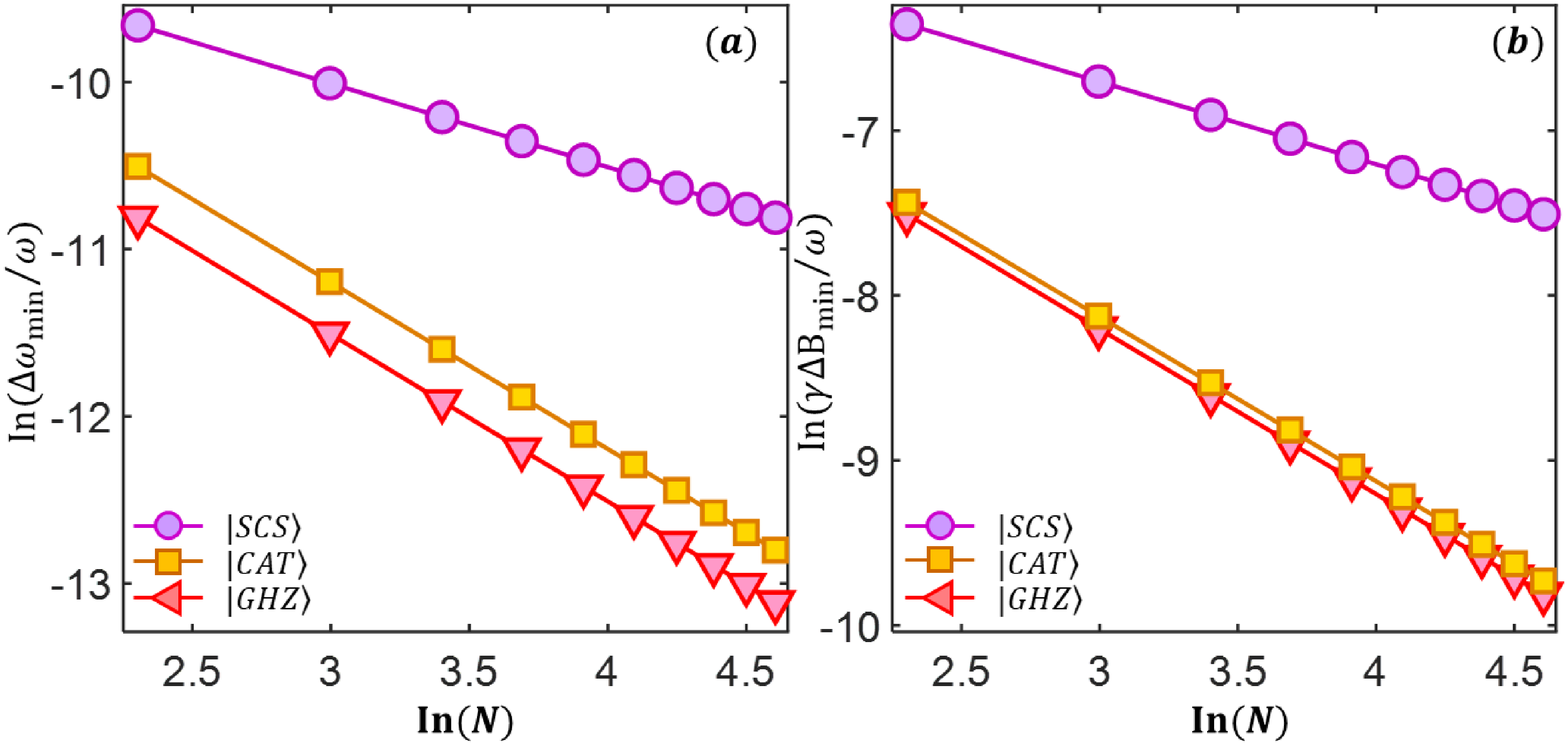}
  \caption{\label{Fig4}(color online).
  Log-log scaling of the optimal measurement precisions (a) $\Delta \omega$ and (b) $\Delta B$ versus total particle number.
  The circles are the results for a SCS with $\hat{U}_{I}=e^{-i\frac{\pi}{2}{\hat{J}_{y}}}$.
  The squares and triangles denote the results for spin cat state $\ket{\Psi(\theta=\pi/8)}_{\textrm{CAT}}$ and GHZ state $\ket{\Psi(\theta=0)}_{\textrm{CAT}}$, respectively, with $\hat{U}_{I}=e^{i\frac{\pi}{2}{\hat{J}_{x}}}e^{i\frac{\pi}{2}{\hat{J}_{z}^2}}e^{i\frac{\pi}{2}{\hat{J}_{x}}}$.
  The lines are the corresponding fitting curves.
  Here, $T=2\pi$, $\omega=20\pi$, $|(\omega-\omega_{e})/\omega|\in[0,0.008]$, and $\gamma B=2\pi$.
  }
\end{figure}
Spin cat states are typical kinds of macroscopic superposition of spin coherent states (MSSCS)~\cite{JHuang2015,JHuang2018012129}.
We consider the states in the form of
\begin{eqnarray}\label{General_MSSCS1}
	\ket{\Psi(\theta)}_{\textrm{MSSCS}}&=&N_c\left(\ket{\Psi(\theta)}_{\textrm{SCS}}+\ket{\Psi(\pi-\theta)}_{\textrm{SCS}}\right)\nonumber \\
                      &=&{N}_c\left(\sum_{m=-J}^{J}\left[c_m^{J}(\theta)+c_{m}^{J}(\pi-\theta)\right]\ket{J,m}\right)\nonumber \\
                      &=&{N}_c \left(\sum_{m=-J}^{J} c_m^{J}(\theta)\left(\ket{J,m} + \ket{J,-m}\right)\right), \nonumber \\
\end{eqnarray}
%
%
%
%
where $\textsl{N}_c$ is the normalization and $\ket{\Psi(\theta)}_{\textrm{SCS}}$ denotes the $N$-particle SCS $\ket{\Psi(\theta)}_{\textrm{SCS}}=\sum_{m=J}^{J} c_{m}^{J}(\theta)\ket{J,m}$ with $c_{m}^{J}(\theta)=\sqrt{\frac{(2J)!}{(J+m)!(J-m)!}}\cos^{J+m}(\frac{\theta}{2})\sin^{J-m}(\frac{\theta}{2})$.
%

%
%
%
%

%
%
For total particle number $N\geq6$, when $\theta \leq \pi/8$, the corresponding MSSCS can be regarded as a spin cat state~\cite{JHuang2018012129}.
%
%
Under this condition, we approximate the spin cat states as
\begin{eqnarray}\label{spin cat states}
	 \ket{\Psi(\theta)}_{\textrm{CAT}}&=&\frac{1}{\sqrt{2}}\left[\sum_{m=-J}^{J}c_m^{J}(\theta)(\ket{J,m}+\ket{J,-m})\right].\nonumber \\
\end{eqnarray}
%
%

It is known that, the interaction-based readout is a powerful technique for achieving Heisenberg limit via spin cat states without single-particle resolved detection~\cite{EDavis2016,TMacri2016,FFrowis2016,Szigeti2017,Nolan2016,JHuang2018012129,Mirkhalaf2018,Anders2018,Burd2019,Linnemann2016,OHosten2016}, which is now feasible in experiments~\cite{Burd2019,Hosten2016}.
Similarly, we adopt an interaction-based operation $\hat{U}_{I}=e^{i\frac{\pi}{2}{\hat{J}_{x}}}e^{i\frac{\pi}{2}{\hat{J}_{z}^2}}e^{i\frac{\pi}{2}{\hat{J}_{x}}}$ in the readout stage.

Therefore, the final state before the half-population difference can be written as
\begin{eqnarray}\label{Evo_Spin_Cat_state}
	|\Psi\rangle_{\text{final}}^{I} = e^{i\frac{\pi}{2}{\hat{J}_{x}}} e^{i\frac{\pi}{2}{\hat{J}_{z}^{2}}} e^{i\frac{\pi}{2}{\hat{J}_{x}}} e^{-i{\phi}} \ket{\Psi(\theta)}_\textrm{CAT}.
\end{eqnarray}
When $N$ is an even number, the final state $|{\Psi}\rangle_\text{final}^{I}$ has analytical form which is written as
\begin{eqnarray}\label{Evo_Spin_cat_state2}
|\Psi\rangle_{\text{final}}^{I}&&=\sum_{m=-J}^{m=J}\frac{c_{m}^{J}(\theta)}{\sqrt{2}}\left[\cos(m\phi)-(-1)^{J-m}\sin(m\phi)\right]  \nonumber \\
&&\times (i)^{(J-m)^2}\ket{J,m}.
\end{eqnarray}
Thus, the expectation of the half-population difference is
\begin{eqnarray}\label{Jz_Spin_cat_state1}
\langle J_{z} \rangle_{\text{f}}=\frac{(-1)^{L+1}}{2}\sum_{m=-J}^{m=J}\!\!(-1)^{J-m}m\left|c_{m}^{J}(\theta)\right|^2 \sin(2m\phi) \nonumber \\
\end{eqnarray}
and its derivative with respect to $\phi$ reads as,
\begin{eqnarray}\label{Jz_Spin_cat_state2}
\frac{d\langle J_{z} \rangle_{\text{f}}}{d\phi}=(-1)^{L+1}\sum_{m=-J}^{m=J}m^2(-1)^{J-m}\left|c_{m}^{J}(\theta)\right|^2 \cos(2m\phi). \nonumber \\
\end{eqnarray}
The expectation of $\hat{J}_{z}^2$ is $\langle \hat{J}_{z}^{2}\rangle_{\text{f}} = \frac{1}{2}\sum_{m=-J}^{m=J}m^2\left|c_{m}^{J}(\theta)\right|^2$.
Correspondingly, the standard deviation of half-population difference is
\begin{widetext}
\begin{eqnarray}\label{Jz_Spin_cat_state3}
(\Delta\hat{J}_{z})_{\text{f}} = \sqrt{\frac{1}{2} \sum_{m=-J}^{m=J}m^2\left|c_{m}^{J}(\theta)\right|^2\!-\!\left[\frac{1}{2} \sum_{m=-J}^{m=J}(-1)^{J\!-\!m} m \left|c_{m}^{J}(\theta)\right|^2 \sin(2m\phi)\!\right]^2}. \nonumber\\
\end{eqnarray}
\end{widetext}
Now, the oscillation of $\langle J_{z} \rangle_{\text{f}}$ becomes related to $2m$.
The form of Eq.~\eqref{Jz_Spin_cat_state1} is also exactly anti-symmetric with respect to detuning $\omega-\omega_{e}$, see the example for spin cat state with $\theta=\pi/8$ in Fig.~\ref{Fig3} (orange dashed line).
Correspondingly, one can determine the zero detuning point $\omega-\omega_{e}=0$ from the symmetry property of the half-population difference.

Especially when $\theta=0$, the spin cat state $\ket{\Psi(0)}_{\textrm{CAT}}=\frac{1}{\sqrt{2}}\left(\ket{J,J} + \ket{J,-J}\right)$ corresponds to the well-known GHZ state.
According to Eq.~\eqref{Evo_Spin_cat_state2} and Eq.~\eqref{Jz_Spin_cat_state1},  the final state $|\Psi\rangle_{\text{final}}^{I}$ becomes
%
%
\begin{eqnarray}\label{Evo_GHZ}
|\Psi\rangle_{\text{final}}^{I}
&=&\frac{1}{\sqrt{2}}\left[\cos(J\phi)-\sin(J\phi)\right]\ket{J,J}\nonumber \\
&+&\frac{1}{\sqrt{2}}\left[\cos(J\phi)+\sin(J\phi)\right]\ket{J,-J}.
\end{eqnarray}
The expectation of the half-population difference is
\begin{eqnarray}\label{Jz_GHZ}
\langle J_{z} \rangle_{\text{f}}=-\frac{N}{2}(-1)^L\textrm{sin}(N\phi).
\end{eqnarray}
Clearly, the main frequency of the oscillation of $\langle J_{z} \rangle_{\text{f}}$ becomes proportional to $N=2J$.
The function of Eq.~\eqref{Jz_GHZ} is anti-symmetric with respect to $\omega-\omega_{e}$, which can be used to locate the zero detuning point $\omega=\omega_e$, as shown in Fig.~\ref{Fig3} (blue lines).
Moreover, the square of half-population difference is independent on the two parameters, i.e., $\langle J_{z}^{2} \rangle_{\text{f}}={J^{2}}=N^2/4$.
According to Eq.~\eqref{Eq:omega uncertainty} and Eq.~\eqref{Eq:B uncertainty}, the measurement precisions for $\omega$ and $B$ can be analytically obtained,
\begin{equation}\label{GHZ_Delta_omega0}
\Delta{\omega}=\frac{\pi}{2\gamma B N }\frac{(\omega-\omega_{e})^2}{\left|G\right|},
\end{equation}
\begin{equation}\label{GHZ_Delta_B0}
\Delta{B}=\frac{\pi}{2\gamma N}\left|\frac{(\omega-\omega_{e})}
{\cos[(\omega-\omega_{e}) T]-1}\right|.
\end{equation}
Similar to the case of SCS, for a fixed $N$, the measurement precision $\Delta \omega$ is dependent on detuning $\omega-\omega_{e}$, the signal amplitude $B$ and evolution time $T$.
When $\omega-\omega_e \rightarrow 0$, $\Delta \omega \rightarrow \frac{\pi}{3\gamma B T^2 N }$ attains the optimal value, which is Heisenberg-limited.
The measurement precision $\Delta B$ is just dependent on detuning $\omega-\omega_{e}$ and evolution time $T$.
When $\omega-\omega_{e} \rightarrow 0$, $\Delta{B}\rightarrow \frac{\pi}{\gamma (\omega-\omega_{e})T^2 N}\rightarrow \infty$, which is diverged.
Also, when $\cos[(\omega-\omega_e)T] + (\omega-\omega_e)T\sin[(\omega-\omega_e)T]=1$, $\Delta B$ attains its optimal.

For other spin cat states, substituting Eq.~\eqref{Jz_Spin_cat_state2} and Eq.~\eqref{Jz_Spin_cat_state3} into Eq.~\eqref{Eq:omega uncertainty} and Eq.~\eqref{Eq:B uncertainty}, one can analytically obtain the measurement precisions $\Delta \omega$ and $\Delta B$.
\begin{widetext}
They are written as
\begin{equation}\label{Delta_omega1}
\Delta{\omega}\!=\!\frac{\tilde{A}\pi}{2\gamma B}\!\frac{(\omega-\omega_{e})^2}{\left|\cos\left[(\omega-\omega_{e}) T\right]-T(\omega-\omega_{e})\sin\left[(\omega-\omega_{e}) T\right]-1\right|},
\end{equation}
\begin{equation}\label{Delta_B1}
\Delta{B}=\frac{\tilde{A}\pi}{2\gamma}\left|\frac{\omega-\omega_{e}}
{\cos\left[(\omega-\omega_{e}) T\right]-1}\right|.
\end{equation}
Here, the coefficient
\begin{equation}\label{Coefficient_A}
\tilde{A}=\frac{\sqrt{\frac{1}{2}\sum_{m=-J}^{m=J}m^2|c_{m}^{J}(\theta)|^2\!-\!\left[\frac{1}{2}\sum_{m=-J}^{m=J}(-1)^{J-m}m|c_{m}^{J}(\theta)|^2 \sin(2m\phi)\right]^2}}{\left|\sum_{m=-J}^{m=J}m^2(-1)^{J-m}|c_{m}^{J}(\theta)|^2 \cos(2m\phi)\right|}.
\end{equation}

When $\omega-\omega_e\rightarrow$, $\phi=\frac{2\gamma B}{\pi}\frac{1-\cos\left[(\omega-\omega_{e}) T\right]}{\omega-\omega_{e}}\rightarrow 0$, $\sin(2m\pi)\approx0$, and $\cos(2m\pi)\approx1$, the measurement precisions $\Delta \omega$ and $\Delta B$ can be simplified as
\begin{eqnarray}\label{Delta_omega2}
\Delta{\omega}|_{\phi=0}&\approx&\frac{\pi}{2\gamma B}\frac{\sqrt{\frac{1}{2}\!\sum_{m=-J}^{m=J}\!m^2|c_{m}^{J}(\theta)|^2}}{\left|\sum_{m=-J}^{m=J}m^2(-1)^{J-m}|c_{m}^{J}(\theta)|^2\right|}\frac{(\omega-\omega_{e})^2}{\left|\cos[(\omega-\omega_{e}) T]-T(\omega-\omega_{e})\sin[(\omega-\omega_{e}) T]-1\right|}\nonumber \\
&=&\tilde{B}(J,\theta) \cdot \frac{\pi}{2\gamma B}\frac{(\omega-\omega_{e})^2}{\left|\cos[(\omega-\omega_{e}) T]-T(\omega-\omega_{e})\sin[(\omega-\omega_{e}) T]-1\right|}=\tilde{B}(J,\theta) \cdot \frac{\pi}{3\gamma B T^2},
\end{eqnarray}
\begin{eqnarray}\label{Delta_B2}
\Delta{B}|_{\phi=0}&\approx&\frac{\sqrt{\frac{1}{2}\!\sum_{m=-J}^{m=J}\!m^2|c_{m}^{J}(\theta)|^2}}{\left|\sum_{m=-J}^{m=J}m^2(-1)^{J-m}|c_{m}^{J}(\theta)|^2\right|}\frac{\pi}{2\gamma}\left|\frac{\omega-\omega_{e}}
{\cos[(\omega-\omega_{e}) T]-1}\right|\nonumber \\
&=&\tilde{B}(J,\theta) \cdot \frac{\pi}{2\gamma}\frac{\omega-\omega_{e}}
{\cos[(\omega-\omega_{e}) T]-1}=\tilde{B}(J,\theta) \cdot \frac{\pi}{\gamma (\omega-\omega_e)T^2}.
\end{eqnarray}

In particular, when $\phi=\frac{2\gamma B}{\pi}\frac{1-\cos\left[(\omega-\omega_{e}) T\right]}{\omega-\omega_{e}}=\frac{\pi}{2}$, $\sin(m\pi)=0$, and $(-1)^{J-m}\cos(m\pi)=1$, the measurement precisions $\Delta \omega$ and $\Delta B$ can be simplified as
\begin{eqnarray}\label{Delta_omega3}
\Delta{\omega}|_{\phi=\frac{\pi}{2}}&=&\frac{\pi}{2\gamma B}\frac{\sqrt{\frac{1}{2}\!\sum_{m=-J}^{m=J}\!m^2|c_{m}^{J}(\theta)|^2}}{\left|\sum_{m=-J}^{m=J}m^2|c_{m}^{J}(\theta)|^2\right|}\frac{(\omega-\omega_{e})^2}{\left|\cos[(\omega-\omega_{e}) T]-T(\omega-\omega_{e})\sin[(\omega-\omega_{e}) T]-1\right|}\nonumber \\
&=& C(\theta) \cdot \frac{\pi}{2\gamma BN}\frac{(\omega-\omega_{e})^2}{\left|\cos[(\omega-\omega_{e}) T]-T(\omega-\omega_{e})\sin[(\omega-\omega_{e}) T]-1\right|},
\end{eqnarray}
\begin{eqnarray}\label{Delta_B3}
\Delta{B}|_{\phi=\frac{\pi}{2}}&=&\frac{\sqrt{\frac{1}{2}\!\sum_{m=-J}^{m=J}\!m^2|c_{m}^{J}(\theta)|^2}}{\left|\sum_{m=-J}^{m=J}m^2|c_{m}^{J}(\theta)|^2\right|}\frac{\pi}{2\gamma}\left|\frac{\omega-\omega_{e}}
{\cos[(\omega-\omega_{e}) T]-1}\right|
=C(\theta) \cdot \frac{\pi}{2\gamma N}\left|\frac{\omega-\omega_{e}}
{\cos[(\omega-\omega_{e}) T]-1}\right|.
\end{eqnarray}
\end{widetext}
The coefficient $C(\theta)=1+\frac{2\tan^2(\theta/2)}{1+\tan^2(\theta/2)}$ is independent on particle number $N$~\cite{JHuang2018012129}.
Still, when $\omega-\omega_e \rightarrow 0$, $\Delta \omega \rightarrow \frac{\pi}{3\gamma B T^2 N }$ is the optimal value.
Under condition $\cos[(\omega-\omega_e)T] + (\omega-\omega_e)T\sin[(\omega-\omega_e)T]=1$, $\Delta B$ attains its optimal.

Further, we numerically find the optimal measurement precisions in the range of $\left|(\omega-\omega_{e})/\omega\right|\in[0,0.008]$ and calculate the corresponding scaling versus particle number for spin cat state $\ket{\Psi(\theta=\pi/8)}_{\textrm{CAT}}$ and GHZ state.
%
%
As shown in Fig.~\ref{Fig4}, for spin cat state $\ket{\Psi(\theta=\pi/8)}_{\textrm{CAT}}$, the optimal log-log measurement precisions $\Delta \omega_{\textrm{min}}$ (yellow squares) and $\Delta B_{\textrm{min}}$ (yellow squares) are $\textrm{ln}\left({{\Delta \omega_{\textrm{min}}}/{\omega}}\right)\approx -\textrm{ln}({N})-8.44$ and $\textrm{ln}\left({\gamma \Delta B_{\textrm{min}}}/{\omega}\right)\approx -\textrm{ln}({N})-5.09$.
For GHZ state, the optimal log-log measurement precisions $\Delta \omega_{\textrm{min}}$ (red triangles) and $\Delta B_{\textrm{min}}$ (red triangles) are $\textrm{ln}\left({{\Delta \omega_{\textrm{min}}}/{\omega}}\right)\approx -\textrm{ln}({N})-8.51$ and $\textrm{ln}\left({\gamma \Delta B_{\textrm{min}}}/{\omega}\right)\approx -\textrm{ln}({N})-5.20$.
It is indicated that, the measurement precisions $\Delta{\omega}$ and $\Delta B$ for spin cat states can exhibit the Heisenberg-limited scaling.

\begin{figure*}[!htp]
 \includegraphics[width=2\columnwidth]{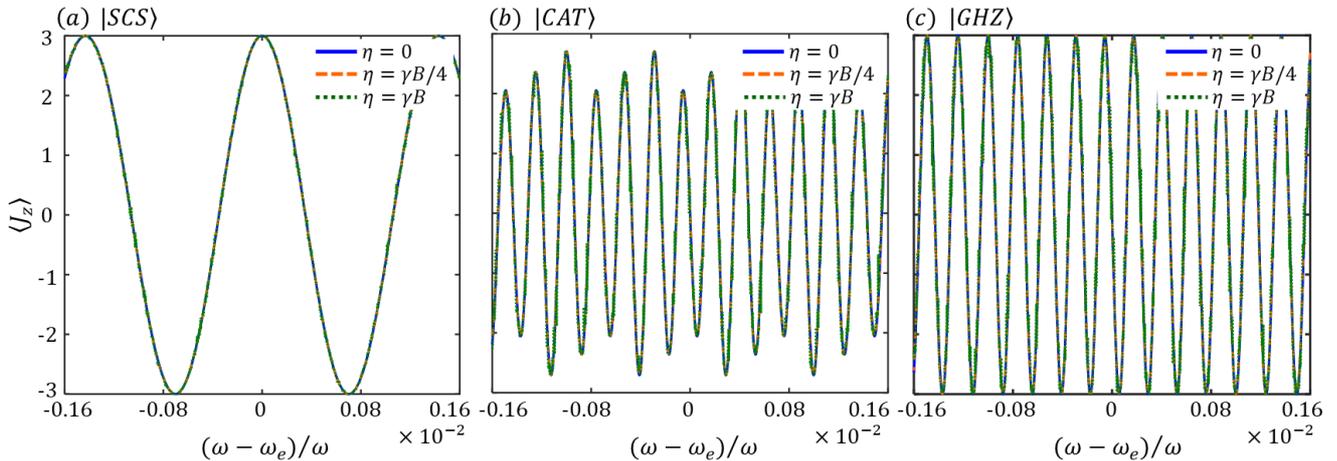}
  \caption{\label{Fig5}(color online).
  The robustness of the many-body quantum amplifiers against stochastic noise. The variation of half-population difference measurement versus detuning for (a) individual particles in SCS, (b) spin cat state $\ket{\Psi(\theta=\pi/8)}_{\textrm{CAT}}$, and GHZ state $\ket{\Psi(\theta=0)}_{\textrm{CAT}}$.
  %
  %
  %
  %
  The blue solid line are the results without noise ($\eta=0$).
  The orange dashed line and green dotted line are the results under noise with $\eta=\gamma B/4$ and $\eta=\gamma B$, repsectively.
  Here, $T=2\pi$, $\omega=20\pi$, $\gamma B=2\pi$, $N=6$, $N_{o}(t)\in [-\eta,\eta]$.
  }
\end{figure*}

\section{robustness\label{Sec5}}
Below, we analyze the robustness of the many-body quantum lock-in amplifier against the stochastic noise.
In realistic experiments, there are many imperfections that limit the parameters estimation.
Here, we will discuss the effect of noise caused by environment on our scheme.

The Hamiltonian for the system coupled to the external AC magnetic field in a noisy environment can be described as
\begin{equation}\label{Eq:Ham_with_noise}
\hat{H}_B=\left[\gamma B\sin(\omega t)+N_{o}(t)\right]\hat{J}_{z}.
\end{equation}
Here, $N_{o}(t)$ is the stochastic noise, $N_{o}(t) \in [-\eta,\eta]$ and its long-time integration $\overline{N_{o}(t)}=0$.
%
%
Due to the modulation term $\hat{H}_\textrm{mix}=\Omega({t})\hat{J}_{x}$ is a set of sharp $\pi$ pulse, we have $\alpha=\int^{t}_{0}\Omega(t')dt'=L\pi$, so that $\cos(\alpha)=\pm1$ and $\sin(\alpha)=0$.
From Eq.~\eqref{Eq:phi1} and Eq.~\eqref{Eq:phi2}, we have
\begin{eqnarray}\label{Eq:phi1DD}
\varphi_1 &=& \int_0^{T}\left[\gamma B\sin(\omega t)+N_{o}(t)\right] \cos(\alpha)dt \nonumber \\
          &=& \int_0^{T}\gamma B\sin(\omega t)\cos(\alpha)dt+\int_0^{t}N_{o}(t)\cos(\alpha)dt,\nonumber \\
\varphi_2 &=& 0.
\end{eqnarray}
Thus, $\int_0^{T}N_{o}(t)\cos(\alpha)dt \approx 0$, $\varphi_1\approx\phi$.
The effect of stochastic noise $N_{o}(t')$ is cancelled out in this integral.
Meanwhile, the contribution of the target signal $S_{0}\sin(\omega t)$ is imprinted into the phase $\phi$ which is sensitively dependent on the detuning $\omega-\omega_e$.
The signal-to-noise ratio can be effectively improved with our many-body quantum amplifier.

To illustrate the robustness of the scheme directly, we numerically calculate the expectation of population versus detuning under stochastic noise $N_{o}(t)$ with $\eta=0$, $\gamma B/4$ and $\gamma B$, see Fig.~\ref{Fig5}.
It is shown that, for modest noise strength $\eta=\gamma B/4$, even for spin cat states, the curves are almost identical to the ones of their corresponding ideal cases.
When the noise strength is extremely strong $\eta=\gamma B$, only small fluctuations appear and the negative impact of the noise can still be eliminated mostly.
Our numerical simulation clearly indicates that, even though extreme stochastic noises exist, our quantum many-body lock-in amplifier can still be used for alternating signal detection with high signal-to-noise ratio.
Moreover, the spin cat states can still be exploited for achieving the Heisenberg-limited quantum lock-in amplifier,  which is feasible in realistic experiments.
\section{summary and discussions\label{Sec6}}
In summary, we have presented a general scheme for realizing a quantum lock-in amplifier via many-body quantum interferometry under suitable periodic modulations.
Based on our protocol, the frequency and amplitude of an alternating signal can be efficiently extracted with high signal-to-noise ratio, even in a noisy environment.
We have analytically studied the measurement precisions for frequency and amplitude of an AC magnetic field via non-entangled and entangled many-body quantum states.
For non-entangled states, the measurement precisions exhibit the SQL scaling as expected.
For entangled particles, by employing the spin cat states as the input states and applying suitable interaction-based readout operations, the measurement precisions can exhibit the Heisenberg-limited scaling.

Moreover, we illustrate the robustness of our many-body quantum amplifier to noise.
In the presence of strong stochastic noise, even for spin cat states, the many-body quantum amplifiers can still be used for AC field sensing.
Our scheme highlights the multi-$\pi$-pulse sequence as a useful technique for quantum sensing and provides a promising method for achieving Heisenberg-limited detection for alternating signals.

In addition, our study may point out a new way for achieving time-dependent signal measurement via many-body quantum lock-in amplifier.
The amplifier demonstrate the potency of quantum lock-in measurement technique, which is readily available for quantum probes.
In single particle systems, the quantum lock-in measurement technique has widely used for frequency metrology~\cite{JMBossl2017}, magnetic field sensing~\cite{JRMaze2008,SKotler2011,SSchmitt2017}, and force detection for the quantum motion of magnetic mechanical resonators~\cite{RShaniv2017}.
With the help of our many-body quantum lock-in amplifier, the measurement precisions for frequency, magnetic field and weak force may be further improved.
This would be beneficial for the development of practical quantum technologies, such as atomic clocks, magnetometers and force detectors.

\acknowledgements{This work is supported by the Key-Area Research and Development Program of GuangDong Province under Grants No. 2019B030330001, the NSFC (Grant No. 11874434, No. 11574405, and No. 11704420), and the Science and Technology Program of Guangzhou (China) under Grants No. 201904020024}
\section*{Supplementary Material}

In Sec.~A, we provide the detailed derivations of the time-evolution in the interaction picture in the main text.
In Sec.~B, we discuss the influence of the pulse number on the operators and their expectations.
In Sec.~C, we confirm the validity of the effective Hamiltonian in the main text.
In Sec.~D, we give detailed results for the variation of the measurement precisions $\Delta \omega$ and $\Delta B$ with detuning $\omega-\omega_{e}$ and amplitude $B$.
In Sec.~E, we discuss the other condition where the initial phase shift between the input and reference signals is different.
%
\begin{widetext}
\subsection{Time-evolution in the interaction picture\label{Sec1}}

In this section, we give the proof of Eq.(2) $\sim$ Eq.(6) in the main text.
The Schr\"{o}dinger equation for an ensemble of two-level bosonic particles evolving under the Hamiltonian ${\hat{H}}=\emph{M}(t)\hat{J}_{z}+\Omega({t})\hat{J}_{x}$ can be written as
\begin{equation}\label{Eq:AppenStatePartial}
i\frac{\partial{\ket{{\Psi}(t)}}}{\partial{t}}=\left[\emph{M}(t)\hat{J}_{z}+\Omega({t})\hat{J}_{x}\right]{\ket{{\Psi}(t)}}.
\end{equation}
For convenience, we move into the interaction picture with respect to ${\hat{H}_\textrm{mix}}=\Omega({t})\hat{J}_{x}$.
%
%
In this case, we have $\ket{\Psi(t)}=e^{-i \int_{0}^{t} {\hat{H}_\textrm{mix}(t')dt'}}\ket{\Psi(t)}_I$, thus the state evolution for $\ket{\Psi(t)}_I$ can be expressed as ($\hbar=1$)
\begin{eqnarray}\label{Eq:AppenStatePartial}
i\frac{\partial{\ket{\Psi(t)}}_I}{\partial{t}}&&=e^{i\int_{0}^{t} {\hat{H}_\textrm{{mix}}(t')dt'}}\left[\emph{M}(t)\hat{J}_{z}\right]e^{-i \int_{0}^{t} {\hat{H}_\textrm{{mix}}(t')dt'}}{\ket{\Psi(t)}_I} \nonumber \\
&&=e^{i \int_{0}^{t} {\Omega({t})\hat{J}_{x}(t')dt'}}\left[\emph{M}(t)\hat{J}_{z}\right]e^{-i \int_{0}^{t} {\Omega({t})\hat{J}_{x}(t')dt'}}{{\ket{\Psi(t)}_I}}
\end{eqnarray}
Defining $\alpha=\int^{t}_{0} \Omega(t')dt'$, and since $[\hat{J}_a, \hat{J}_b]=i\epsilon_{abc}\hat{J}_c$ ($\epsilon_{abc}$ is the Levi-Civita symbol with $a,b,c=x,y,z$), $e^{i\beta\hat{J}_{x}}\hat{J}_{z}e^{-i\beta\hat{J}_{x}}=\cos(\beta)\hat{J}_{z}+\sin(\beta)\hat{J}_{y}$, we have
\begin{equation}\label{Eq:AppenStatePartial}
i\frac{\partial{\ket{\Psi(t)}_I}}{\partial{t}}=\emph{M}(t)\left[\cos(\alpha)\hat{J}_{z}+\sin(\alpha)\hat{J}_{y}\right]{\ket{\Psi(t)}_I}.
\end{equation}
After integration, at time $T$, the evolved state $|\Psi(T)\rangle_I$ becomes
\begin{eqnarray}\label{Eq:AppenStateEvolve}
|\Psi(T)\rangle_I=\hat{\mathcal{T}} e^{-i\int_0^{T}\left[\emph{M}(t)(\cos(\alpha)\hat{J}_{z}+\sin(\alpha)\hat{J}_{y})\right]dt}|{{\Psi}}(0)\rangle_I
= \hat{\mathcal{T}} e^{\hat{A}}|{{\Psi}}(0)\rangle_I.
\end{eqnarray}
Here, $\hat{\mathcal{T}}$ is the time-ordering operator and operator $\hat{A}$ can be written as
\begin{eqnarray}\label{Eq:AppenA}
\hat A&=&-{i\int_0^{T}[\emph{M}(t)\textrm{cos}(\alpha)\hat{J}_{z}+\emph{M}(t)\textrm{sin}(\alpha)\hat{J}_{y}]dt}\nonumber \\
 &=&-{i\int_0^{T}[\emph{M}(t)\textrm{cos}(\alpha)dt]\hat{J}_{z}}+{i\int_0^{T}[\emph{M}(t)\textrm{sin}(\alpha)dt]\hat{J}_{y}}\nonumber \\
 &=&-{i(\varphi_1\hat{J}_{z}}+{\varphi_2\hat{J}_{y}}).
\end{eqnarray}
Thus, we have
\begin{eqnarray}\label{Eq:StateEvolve}
|\Psi(T)\rangle_I =\hat{\mathcal{T}} e^{-i({\varphi_1\hat{J}_{z}}+{\varphi_2\hat{J}_{y}})}|{{\Psi}}(0)\rangle_{I},
\end{eqnarray}
with $|{{\Psi}}(0)\rangle_{I}=|{{\Psi}}(0)\rangle$ and the two phase factors
\begin{eqnarray}\label{Eq:phi1}
\varphi_1= \int_0^{T}\emph{M}(t)\cos(\alpha)dt
         = \int_0^{T}\left[S_{0}\sin(\omega t+\beta)+\emph{N}_{o}(t)\right]\cos(\alpha)dt,
\end{eqnarray}
and %
\begin{eqnarray}\label{Eq:phi2}
\varphi_2 = \int_0^{T}\emph{M}(t)\sin(\alpha)dt
          = \int_0^{T}\left[S_{0}\sin(\omega t+\beta)+\emph{N}_{o}(t)\right]\sin(\alpha)dt.
\end{eqnarray}
If one adding the suitable modulation $\Omega(t)$ to make $\textrm{cos}(\alpha)$ and $\textrm{sin}(\alpha)$ are periodic and synchronized with single $\emph{S}(t)$, the phase accumulated owing to $\emph{S}(t)$ can adds up coherently whereas the phase accumulated owing to $\emph{N}_{o}(t)$ can averaged away.
%

\subsection{Influence of pulse number\label{Sec2}}
Here, we discuss the influence of the pulse number in our scheme.
Since we work in the interaction picture, the operator can be written as,
\begin{eqnarray}\label{Observable}
\hat{O}^{{I}}(t) =e^{i\alpha\hat{J}_{x}} \hat{O} e^{-i\alpha\hat{J}_{x}},
\end{eqnarray}
where $\hat{O}$ is the operator in Schr\"{o}dinger picture.
In our scheme, we consider the modulation is a set of sharp $\pi$ pulses and can be approximated as $\Omega(t)=\pi\sum_{m=1}^{M} \delta \left(t-(m+\frac{1}{2})\tau \right)$.
Thus, at the time $T$, the operator in the interaction picture can be written as,
\begin{eqnarray}\label{Observable1}
\hat{O}^{{I}}(T)&& =e^{i\int^{T}_{0} \pi\sum_{m=1}^{M} \delta \left(t-(m+\frac{1}{2})\tau \right) dt\hat{J}_{x}} \hat{O} e^{-i\int^{T}_{0} \pi\sum_{m=1}^{M} \delta \left(t-(m+\frac{1}{2})\tau \right) dt\hat{J}_{x}}.
\end{eqnarray}
When the evolution time $T$ is fixed, for the $i$-th pulse spacing $\tau_i$, the number of the $\pi$ pulses is $L = [T/\tau_i+1/2]$, where $[\cdot]$ stands for the integer-valued division.
Due to the property of the delta functions, we have $\int^{T}_{0} \pi\sum_{m=1}^{M} \delta \left(t-(m+\frac{1}{2})\tau \right) dt = L \pi$.
Therefore, the operators in the interaction picture becomes
\begin{eqnarray}\label{Observable2}
\hat{O}^{{I}}(t)&& =e^{iL\pi\hat{J}_{x}} \hat{O} e^{-iL\pi\hat{J}_{x}}.
\end{eqnarray}
For the collective spin operators, we have
\begin{eqnarray}\label{Observable2}
&&\hat{J}_{x}^{{I}}=e^{iL\pi\hat{J}_{x}} \hat{J}_{x} e^{-iL\pi\hat{J}_{x}}=\hat{J}_{x},\nonumber\\
&&\hat{J}_{y}^{{I}}=e^{iL\pi\hat{J}_{x}} \hat{J}_{y} e^{-iL\pi\hat{J}_{x}}=\cos(L\pi)\hat{J}_{y}-\sin(L\pi)\hat{J}_{z},\nonumber\\
&&\hat{J}_{z}^{{I}}=e^{iL\pi\hat{J}_{x}} \hat{J}_{z} e^{-iL\pi\hat{J}_{x}}=\cos(L\pi)\hat{J}_{z}+\sin(L\pi)\hat{J}_{y}.
\end{eqnarray}
According to Eqs.~\eqref{Observable2}, we find that the parity of pulse number alters the operators.
When $L$ is an even number, the collective spin operators in the interaction picture are
\begin{eqnarray}\label{Observable2}
&&\hat{J}_{x}^{{I}}=\hat{J}_{x},\nonumber\\
&&\hat{J}_{y}^{{I}}=\hat{J}_{y},\nonumber\\
&&\hat{J}_{z}^{{I}}=\hat{J}_{z}.
\end{eqnarray}
While $L$ is an odd number, the operators become
\begin{eqnarray}\label{Observable2}
&&\hat{J}_{x}^{\textrm{I}}=\hat{J}_{x},\nonumber\\
&&\hat{J}_{y}^{\textrm{I}}=-\hat{J}_{y},\nonumber\\
&&\hat{J}_{z}^{\textrm{I}}=-\hat{J}_{z}.
\end{eqnarray}
Except for $\hat{J}_{x}^{\textrm{I}}$, the operators $\hat{J}_{y}^{\textrm{I}}$ and $\hat{J}_{z}^{\textrm{I}}$ change the sign when $L$ is odd.
Thus, the half-population difference of the final state $\langle \hat{J}_z\rangle_{\textrm{f}}$ in Eqs.~(16), (27), (31) are dependent on the parity of pulse number $L$.
%
%
\subsection{Validity of the effective Hamiltonian\label{Sec3}}
%
In the interaction picture, the Hamiltonian for the interrogation stage reads as
\begin{equation}\label{Eq:Hams1}
\hat{H}_\textrm{I}=h(t)[\gamma B \sin(\omega t)]\hat{J}_{z},
\end{equation}
where $h(t)$ is the square wave function.
When $\Omega(t)=\pi\sum_{m=1}^{M} \delta \left(t-(m+\frac{1}{2})\tau \right)$ and the corresponding square wave function is
$$ h(t)=\left\{
\begin{array}{rcl}
1  &&{\omega_{e}t+\pi/2}\quad {\textrm{mod}\quad 2\pi\in[0,\pi)}\\
-1 &&{\omega_{e}t+\pi/2}\quad{ \textrm{mod}\quad 2\pi\in[\pi,2\pi)}
\end{array} \right. $$
Under the condition $|\omega-\omega_{e}|\ll\omega$ and $T\gg2\pi/\omega$, the Hamiltonian~\eqref{Eq:Hams1} can be approximated by an effective Hamiltonian~\cite{Nature47361,Science356832}
\begin{equation}\label{Eq:Hameff1}
\hat{H}_I^{{\textrm{eff}}}=\frac{2 \gamma B}{\pi}\sin[(\omega-\omega_{e}) t]\hat{J}_{z}.
\end{equation}
%
%
%
To compare with the two Hamiltonians, we denote the integrals over time as
\begin{eqnarray}\label{Eq:1}
\int_{0}^{T}\hat{H}_\textrm{I}&=&\hat{J}_{z}\Phi_{1},\nonumber\\
\int_{0}^{T}\hat{H}_\textrm{I}^{{\textrm{eff}}}&=&\hat{J}_{z}\Phi_{2}.
\end{eqnarray}
Here,  $\Phi_{1}$ and $\Phi_{2}$ are two phases dependent on frequency $\omega$ and amplitude $B$, which can be expressed as
\begin{eqnarray}\label{Eq:1}
\Phi_{1}&=&\int_{0}^{T}h(t)[ \gamma B \textrm{sin}(\omega t)]dt,\nonumber\\
\Phi_{2}&=&\int_{0}^{T}\frac{2 \gamma B}{\pi}\textrm{sin}[(\omega-\omega_{e}) t]dt.
\end{eqnarray}
Here, we confirm the validity of the effective Hamiltonian~\eqref{Eq:Hameff1} via numerical calculations.
According to our numerical results, we find that the two phases is almost the same when $\frac{|\omega-\omega_{e}|}{\omega}\leq0.1$, i.e., $\Phi_{1}\approx\Phi_{2}$, for evolution time $T=2\pi$, $\omega\geq10\pi$, as shown in Fig~\ref{FigS1}.
\begin{figure}[!htp]
 \includegraphics[width=0.75\columnwidth]{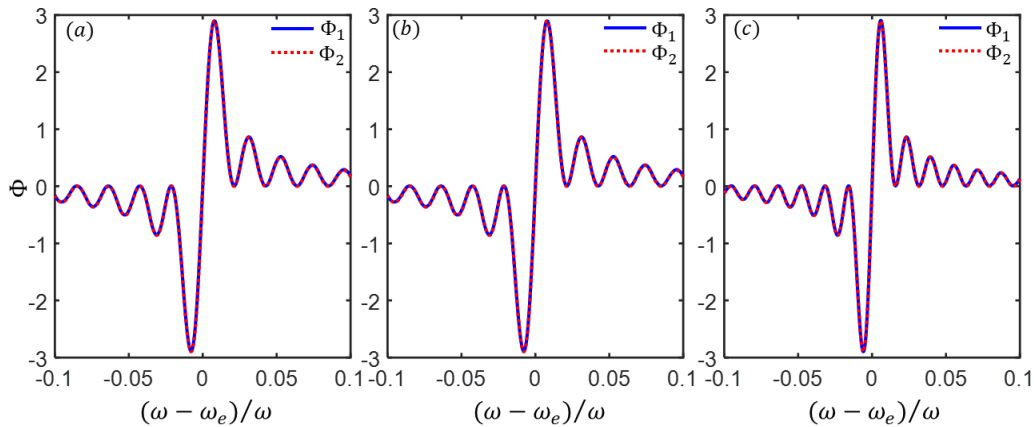}
  \caption{\label{FigS1}(color online).
  The variations of the two phases $\Phi_{1}$ and $\Phi_{2}$ versus $\omega-\omega_{e}$ for (a)$\omega=10 \pi$, (b)$\omega=15\pi$, (c)$\omega=20\pi$.
  Here, $T=2\pi$, $\gamma B=1$.
  }
\end{figure}
%
%
\subsection{Variation of the measurement precison\label{Sec4}}
In the main text, we only show the optimal measurement precision for the two parameters.
In this section, we will give the variation of the measurement precision with detuning $\omega-\omega_{e}$ and amplitude $B$ in detail.
According to the analytical expression of the measurement precision, we find that the measurement precisions $\Delta \omega$ and $\Delta B$ are dependent on both the frequency $\omega$ and amplitude $B$ simultaneously.
For comparison, we choose three typical input states: spin coherent state (SCS) $|\Psi\rangle_{\textrm{SCS}}$, spin cat state $|\Psi(\pi/8)\rangle_{\textrm{CAT}}$, and GHZ state $|\Psi(0)\rangle_{\textrm{CAT}}$.
In Fig.~\ref{FigS2}, the variations of $\Delta \omega/\omega$ and $\gamma \Delta B/\omega$ versus $B$ and $\omega-\omega_e$ with the three input states are shown.
When the input state is a SCS or a GHZ state, the measurement precision $\Delta \omega$ attains its optimal value when $\cos[(\omega-\omega_{e})T][2-(\omega-\omega_{e})^2T^{2}]+2T(\omega-\omega_{e})\sin[(\omega-\omega_{e})T]=2$ under a fixed particle number $N$.
Meanwhile, the measurement precision $\Delta B$ is just dependent on the frequency $\omega$.
The measurement precision of $\Delta{B}$ attains its optimal value when $\cos[(\omega-\omega_{e})T]+{(\omega-\omega_{e}){T}\sin[(\omega-\omega_{e})T]}=1$.
In contrast, the dependence of $\Delta \omega/\omega$ and $\gamma \Delta B/\omega$ on $B$ and $\omega-\omega_e$ with spin cat states $|\Psi(\theta)\rangle_{\textrm{CAT}}$ ($\theta\neq0$) is more complicated, e.g. see Fig.~\ref{FigS2} (b) and (e).
%
\begin{figure}[!htp]
 \includegraphics[width=0.75\columnwidth]{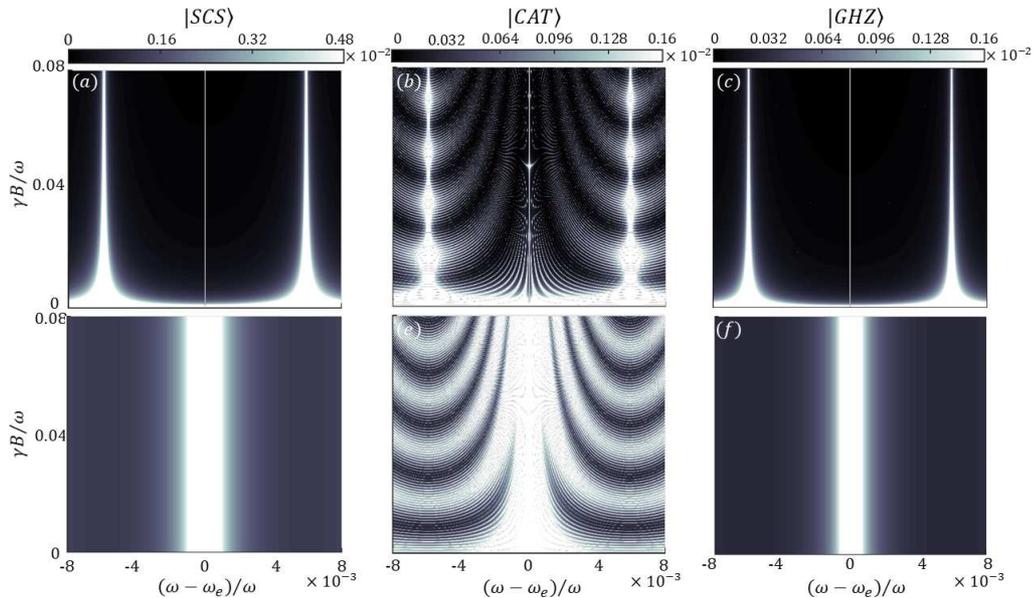}
  \caption{\label{FigS2}(color online).
  The variations of measurement precisions for individual particles and entangled particles under fixed $N$.
  (a) $\Delta \omega/\omega$, (d) $\gamma \Delta B/\omega$ for a spin coherent state (SCS) with $\hat{U}_{I}=e^{-i\frac{\pi}{2}{\hat{J}_{y}}}$.
  (b) $\Delta \omega/\omega$, (e) $\gamma \Delta B/\omega$ for a spin cat state with $\hat{U}_{I}=e^{i\frac{\pi}{2}{\hat{J}_{x}}}e^{i\frac{\pi}{2}{\hat{J}_{z}^2}}e^{i\frac{\pi}{2}{\hat{J}_{x}}}$ and $\theta=\pi/8$.
  (c) $\Delta \omega/\omega$, (f) $\gamma \Delta B/\omega$ for a GHZ state with $\hat{U}_{I}=e^{i\frac{\pi}{2}{\hat{J}_{x}}}e^{i\frac{\pi}{2}{\hat{J}_{z}^2}}e^{i\frac{\pi}{2}{\hat{J}_{x}}}$.
  Here, $T=2\pi$, $\omega=20\pi$, $N=20$.
  }
\end{figure}

\subsection{Influence of initial phase shift between the input and reference signals\label{Sec5}}
In our scheme, the modulation of $\Omega(t)$ is a set of $\pi$-pulse in the form of $\Omega(t)=\pi\sum_{m=1}^{M} \delta \left(t-(m+\frac{1+k}{2})\tau \right)$ where $k\in[-1,1)$.
The parameter $k$ corresponds to the initial phase shift between the input and the reference signals.
The lock-in case of $k=0$ has been discussed.
In the following, we consider the other condition where $k=-1$.

The Hamiltonian for the interrogation stage (after moving to the interaction picture with respect to the modulation) reads
\begin{equation}\label{Eq:Ham1}
\hat{H}_\textrm{I}=h(t)[ \gamma B \sin(\omega t)]\hat{J}_{z}.
\end{equation}
where the square wave function $h(t)$ becomes
$$ h(t)=\left\{
\begin{array}{rcl}
1  &&{\omega_{e}t}\quad {\textrm{mod}\quad 2\pi\in[0,\pi)}\\
-1  &&{\omega_{e}t} \quad{ \textrm{mod}\quad 2\pi\in[\pi,2\pi)}
\end{array} \right. $$
Similarly,  one can easily tune $\omega_e$ close to $\omega$ in experiment.
When $|\omega-\omega_{e}|\ll\omega$ and $T\gg2\pi/\omega$, the Hamiltonian~\eqref{Eq:Ham1} can be approximated by an effective Hamiltonian (the approximation is also valid for $|\omega-\omega_{e}|/\omega < 0.1$ when $\omega>10\pi$),
%
%
\begin{equation}\label{Eq:Hameff1}
\hat{H}_{I}^{\textrm{eff}}=\frac{2 \gamma B}{\pi}\cos\left[(\omega-\omega_{e}) t\right]\hat{J}_{z}.
\end{equation}
%
%
%
%
After similar procedures with the case of $k=1$, the final state before half-population difference measurement can be written as
\begin{eqnarray}\label{Eq:final_state1}
	\ket{\Psi}_{\textrm{final}}=\hat{U}e^{-i\alpha\hat{J}_x}e^{-i\phi_{1}\hat{J}_z} \ket{\Psi}_{\textrm{in}}.
\end{eqnarray}
The analytical form of the phase is written as
\begin{equation}\label{Eq:phase1}
	\phi_{1}=\frac{2 \gamma B}{\pi}\frac{\sin[(\omega-\omega_{e}) T]-1}{\omega-\omega_{e}}.
\end{equation}
\begin{figure}[!htp]
 \includegraphics[width=0.75\columnwidth]{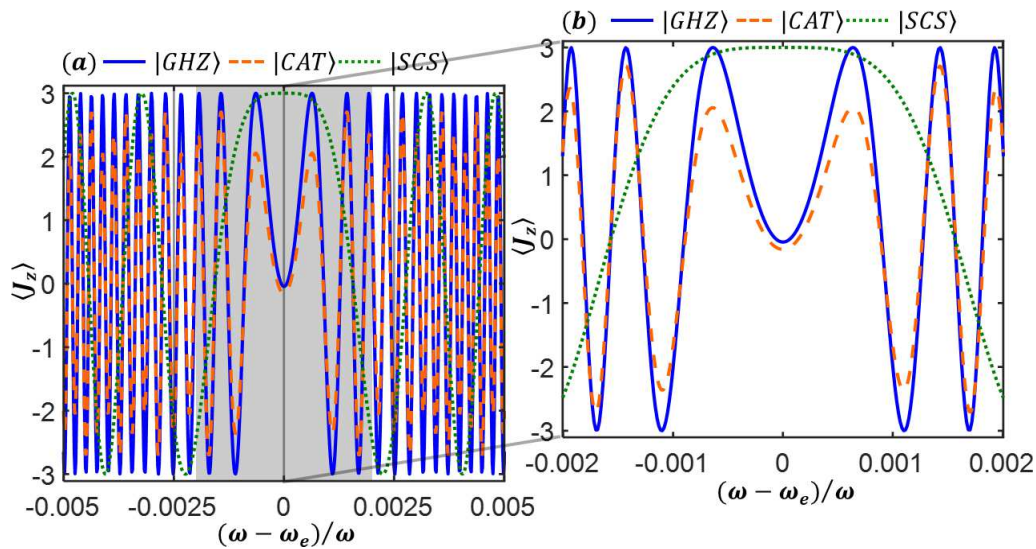}
  \caption{\label{FigS3}(color online).
  The variation of half-population difference measurement versus detuning $\omega-\omega_{e}$ for individual particles and entangled particles under fixed $B$.
  The green dot line show results for a SCS with $\hat{U}_{I}=e^{-i\frac{\pi}{2}{\hat{J}_{y}}}$.
  The orange dashed line show results for a spin cat state with $\hat{U}_{I}=e^{i\frac{\pi}{2}{\hat{J}_{x}}}e^{i\frac{\pi}{2}{\hat{J}_{z}^2}}e^{i\frac{\pi}{2}{\hat{J}_{x}}}$ and $\theta=\pi/8$.
  The blue solid line show results for a GHZ state with $\hat{U}_{I}=e^{i\frac{\pi}{2}{\hat{J}_{x}}}e^{i\frac{\pi}{2}{\hat{J}_{z}^2}}e^{i\frac{\pi}{2}{\hat{J}_{x}}}$.
  These results indicates that the expectation of half-population difference is symmetry respective detuning $\omega-\omega_{e}$ and the symmetry point located at $\omega-\omega_{e}=0$, i.e., $\omega=\omega_{e}$.
  Here, $T=2\pi$, $\omega=20\pi$, $\gamma B=2\pi$, $N=6$ and $k=0$.
  }
\end{figure}
\begin{figure}[!htp]
 \includegraphics[width=0.75\columnwidth]{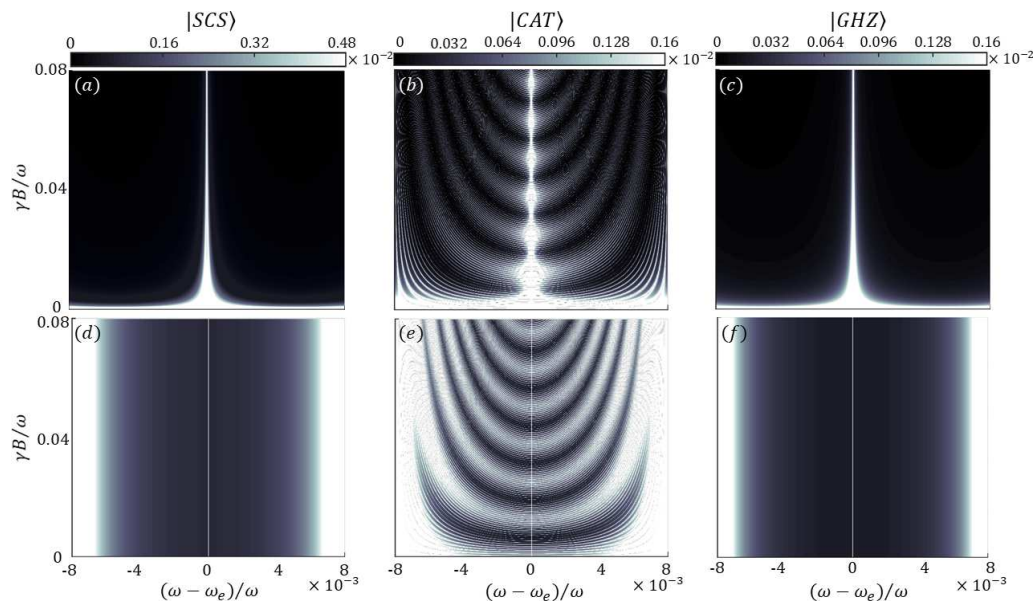}
  \caption{\label{FigS4}(color online).
  The variations of measurement precisions for individual particles and entangled particles under fixed $N$.
  (a) $\Delta \omega/\omega$,(d)$\gamma \Delta B/\omega$ for a SCS with $\hat{U}_{I}=e^{-i\frac{\pi}{2}{\hat{J}_{y}}}$.
  (b)$\Delta \omega/\omega$,(e)$\gamma \Delta B/\omega$ for a spin cat state with $\hat{U}_{I}=e^{i\frac{\pi}{2}{\hat{J}_{x}}}e^{i\frac{\pi}{2}{\hat{J}_{z}^2}}e^{i\frac{\pi}{2}{\hat{J}_{x}}}$ and $\theta=\pi/8$.
  (c) $\Delta \omega/\omega$,(f) $\gamma \Delta B/\omega$ for a GHZ state with $\hat{U}_{I}=e^{i\frac{\pi}{2}{\hat{J}_{x}}}e^{i\frac{\pi}{2}{\hat{J}_{z}^2}}e^{i\frac{\pi}{2}{\hat{J}_{x}}}$.
  Here, $T=2\pi$, $\omega=20\pi$, $N=20$ and $k=0$.
  }
\end{figure}

Still, we first consider individual particles without any entanglement and then discuss the cases of spin cat states.
%
%
%
%
%
For spin coherent state $\ket{\Psi}_{\textrm{SCS}}$, the final state before the half-population difference measurement can be written as
\begin{eqnarray}\label{Evo_CSC}
|\Psi\rangle_{\text{final}}^{I}
=e^{-i\frac{\pi}{2}{\hat{J}_{y}}} e^{-i \phi_{1} \hat J_z}\ket{\Psi}_{\textrm{SCS}}.
\end{eqnarray}
In an explicit form, the final state becomes
\begin{eqnarray}\label{Evo_CSC}
|\Psi\rangle_{\text{final}}^{I}=\!\!\!\sum_{m=\!-J}^{J}\!\!\!\sqrt{C_{J}^{m}}\left[-\cos(\frac{\phi_{1}}{2})\right]^{\!J\!+\!m}\left[i\sin(\frac{\phi_{1}}{2})\right]^{\!J\!-\!m}\!\!\!|J,m\rangle. \nonumber \\
\end{eqnarray}
where $C_{J}^{m}={\frac{(2J)!}{(J+m)!(J-m)!}}$ is the binomial coefficient.
%
%
After some algebra, the expectations of half-population difference and the square of half-population difference on the final state can also be explicitly written as

\begin{eqnarray}\label{Jz_SCS}
\langle J_{z} \rangle_{\text{f}} = (-1)^L \frac{N}{2}\cos(\phi_{1})
\end{eqnarray}
%
%
\begin{eqnarray}\label{Jz2_SCS}
\langle J_{z}^{2} \rangle_{\text{f}} =\frac{N}{4}+\frac{N(N-1)}{4}\cos^{2}(\phi_{1})
\end{eqnarray}
From Eq.~\eqref{Jz_SCS}, it is found that the information of the estimated two parameters $\omega$ and $B$ can be inferred from the bi-sinusoidal oscillation of the half-population difference.
In our calculation, we find the function of Eq.~\eqref{Jz_SCS} with respect to detuning $\omega-\omega_{e}$ is symmetry under fixed $B$ and the symmetry point located at $\omega-\omega_{e}=0$, i.e., $\omega=\omega_e$, as shown in Fig.~\ref{FigS3} (green dot line).
Thus, one can obtain the value of $\omega$, and further extract the value of $B$ according to the analytic form of Eq.~\eqref{Jz_SCS}.
This imply that the values of $\omega$ and $B$ can be simultaneously obtained only by the half-population difference measurement.

%

According to the quantum estimation theory~\cite{Helstrom1967,CWHelstrom1976,Paris2009,AdvPhysX2016}, the measurement precisions of the estimated parameters can be given according to the error propagation formula.
The measurement precisions for $\omega$ and $B$ are
\begin{equation}\label{Eq:omega uncertainty}
\Delta \omega=\frac{(\Delta{\hat{J}_{z}})_{\text{f}}}{|\partial{\langle\hat{J}_{z}\rangle_{\text{f}}}/ \partial{\omega_{e}}|},
\end{equation}
and
\begin{equation}\label{Eq:B uncertainty}
\Delta B=\frac{(\Delta{\hat{J}_{z}})_{\text{f}}}{|\partial{\langle\hat{J}_{z}\rangle_{\text{f}}}/ \partial{B}|},
\end{equation}
where
\begin{equation}\label{Eq:Deviation}
(\Delta{\hat{J}_{z}})_{\text{f}}=\sqrt{\langle\hat{J}_z^2\rangle_{\text{f}}-\langle\hat{J}_{z}\rangle_{\text{f}}^2}.
\end{equation}

Substituting Eq.~\eqref{Jz_SCS} and Eq.~\eqref{Jz2_SCS} into Eq.~\eqref{Eq:omega uncertainty} and Eq.~\eqref{Eq:B uncertainty}, one can analytically obtain the measurement precisions $\Delta \omega$ and $\Delta B$.
They read
\begin{equation}\label{SCS_Delta_omega0}
\Delta{\omega}\!=\!\frac{\pi}{2\gamma B\sqrt{N}}\!\frac{(\omega-\omega_{e})^2}{\left|\sin\left[(\omega-\omega_{e}) T\right]-T(\omega-\omega_{e})\cos\left[(\omega-\omega_{e}) T\right]\right|},
\end{equation}
\begin{equation}\label{SCS_Delta_B0}
\Delta{B}=\frac{\pi}{2\gamma\sqrt{N}}\left|\frac{\omega-\omega_{e}}
{\sin\left[(\omega-\omega_{e}) T\right]}\right|.
\end{equation}
From Eq.~\eqref{SCS_Delta_omega0} and Eq.~\eqref{SCS_Delta_B0}, the measurement precisions $\Delta{\omega}$ and $\Delta B$  only exhibit the SQL scaling.
For a fixed $N$, the measurement precision $\Delta \omega$ is dependent on both detuning $\omega-\omega_{e}$ and $B$, as shown in Fig.~\ref{FigS4}~(a).
When $\sin[(\omega-\omega_{e})T](2-(\omega-\omega_{e})^2T^{2})-2T(\omega-\omega_{e})\cos[(\omega-\omega_{e})T]=0$, $\Delta \omega$ attains the optimal value.
However, the measurement precision $\Delta B$ is just dependent on detuning $\omega-\omega_{e}$, as shown in Fig.~\ref{FigS4}~(d).
When it is near resonance $\omega\approx\omega_{e}$, $\Delta{B}=\frac{\pi}{2\gamma\sqrt{N}T}$ is optimal.

We numerically calculate the optimal measurement precisions versus particle number in the range of $\left|(\omega-\omega_{e})/\omega\right|\in[0,0.008]$ and $\gamma B=2\pi$.
According to fitting results (see Fig.~\ref{FigS5}), the log-log measurement precision for $\textrm{ln}\left({{\Delta \omega_{\textrm{min}}}/{\omega}}\right)\approx -0.5\textrm{ln}({N})-8.37$ and $\textrm{ln}\left({\gamma \Delta B_{\textrm{min}}}/{\omega}\right)\approx -0.5\textrm{ln}({N})-5.53$.
For the two parameters, the optimal measurement precision $\Delta \omega_{\textrm{min}}$(purple circles) and $\Delta B_{\textrm{min}}$ (purple circles) are both exhibit SQL scaling as expected.
%

%
\begin{figure}[!htp]
 \includegraphics[width=0.75\columnwidth]{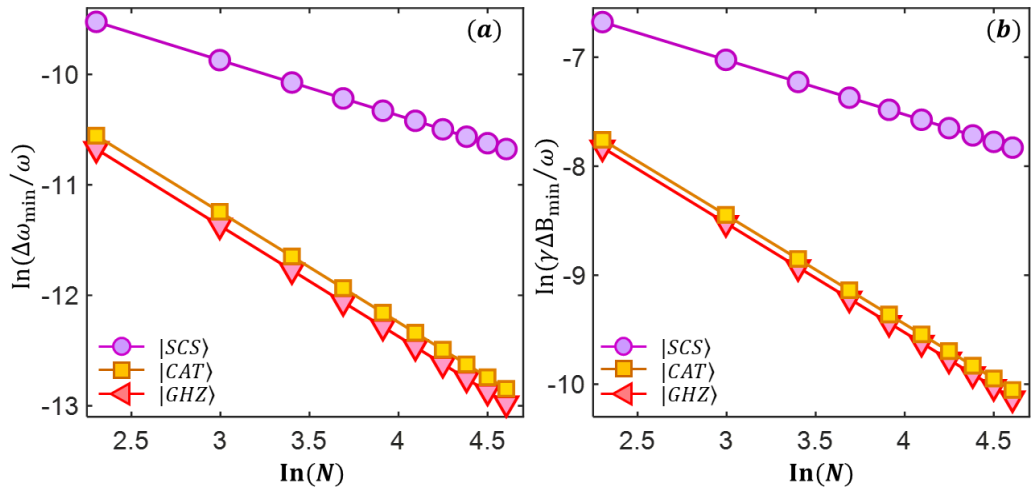}
  \caption{\label{FigS5}(color online).
  Log-log plots of the scaling of the optimal measurement precision versus total particle number for (a) frequency $\omega$ and (b) amplitude $B$.
  The circles show results for a SCS with $\hat{U}_{I}=e^{-i\frac{\pi}{2}{\hat{J}_{y}}}$ and the lines are the corresponding fitting curves.
  The square show results for a spin cat state with $\hat{U}_{I}=e^{i\frac{\pi}{2}{\hat{J}_{x}}}e^{i\frac{\pi}{2}{\hat{J}_{z}^2}}e^{i\frac{\pi}{2}{\hat{J}_{x}}}$ and $\theta=\pi/8$, the lines are the corresponding fitting curves.
  The triangle show results for a GHZ state with $\hat{U}_{I}=e^{i\frac{\pi}{2}{\hat{J}_{x}}}e^{i\frac{\pi}{2}{\hat{J}_{z}^2}}e^{i\frac{\pi}{2}{\hat{J}_{x}}}$, the lines are the corresponding fitting curves.
  Here, $T=2\pi$, $\omega=20\pi$, $|(\omega-\omega_{e})/\omega|\in[0,0.008]$, $\gamma B\in[0,5]$ and $k=-1$.
  }
\end{figure}

Next, we try to use the spin cat states to perform the measurement.
We choose an interaction-based operation $\hat{U}_{I}=\hat{U}=e^{i\frac{\pi}{2}{\hat{J}_{x}}}e^{i\frac{\pi}{2}{\hat{J}_{z}^2}}e^{i\frac{\pi}{2}{\hat{J}_{x}}}$ in the readout stage.
Therefore, the final state before the half-population difference can be written as
\begin{eqnarray}\label{Evo_Spin_Cat_state}
	|\Psi\rangle_{\text{final}}^{I} = e^{i\frac{\pi}{2}{\hat{J}_{x}}} e^{i\frac{\pi}{2}{\hat{J}_{z}^{2}}} e^{i\frac{\pi}{2}{\hat{J}_{x}}} e^{-i{\phi}_{1}{\hat{J}_{z}}} \ket{\Psi(\theta)}_\textrm{CAT}.
\end{eqnarray}
The final state $|\Psi_{\text{final}}\rangle$ has analytical form when $N$ is an even number and can be written as
\begin{eqnarray}\label{Evo_Spin_cat_state}
|\Psi\rangle_{\text{final}}^{I}&&=\sum_{m=-J}^{m=J}\frac{c_{m}^{J}(\theta)}{\sqrt{2}}[\cos(m\phi_1)-(-1)^{(J-m)}\sin(m\phi_1)]  \nonumber \\
&&\times (i)^{(J-m)^2}\ket{J,m}.
\end{eqnarray}
Thus, the expectation of the half-population difference and the square of half-population difference are
\begin{eqnarray}\label{Jz_Spin_cat_state}
\langle J_{z} \rangle_{\text{f}}&=&\frac{(-1)^{L+1}}{2}\sum_{m=-J}^{m=J}\!\!(-1)^{J-m}m|c_{m}^{J}(\theta)|^2 \textrm{sin}(2m\phi_1), \nonumber \\
\langle J_{z}^2 \rangle_{\text{f}}&=&\frac{1}{2}\sum_{m=-J}^{m=J}m^2|c_{m}^{J}(\theta)|^2.
\end{eqnarray}
Its derivative with respect to $\phi_1$ reads as
\begin{eqnarray}\label{Jz_Spin_cat_state}
\frac{\langle dJ_{z} \rangle_{\text{f}}}{d\phi_1}=(-1)^{L+1}\sum_{m=-J}^{m=J}m^2(-1)^{J-m}|c_{m}^{J}(\theta)|^2 \cos(2m\phi_1).
\end{eqnarray}
Correspondingly, the standard deviation of half-population difference is

\begin{eqnarray}\label{Jz_Spin_cat_state}
(\Delta\hat{J}_{z})_{\text{f}}=\sqrt{\langle J_{z}^{2} \rangle_{\text{f}}-(\langle J_{z} \rangle_{\text{f}})^2}
=\sqrt{\frac{1}{2}\!\sum_{m=-J}^{m=J}\!m^2|c_{m}^{J}(\theta)|^2\!-\!\left[\frac{1}{2}\!\sum_{m=-J}^{m=J}\!\!(\!-1\!)^{J\!-\!m}m|c_{m}^{J}(\theta)|^2 \textrm{sin}(2m\phi_1)\right]^2} \nonumber\\
\end{eqnarray}
Now, the main frequencies of the bi-sinusoidal oscillation of $\langle J_{z} \rangle_{\text{f}}$ becomes proportional to $2m$.
Further, we find the function of Eq.~\eqref{Jz_Spin_cat_state} with respect to detuning $\omega-\omega_e$ is symmetry under fixed $B$ and the symmetry point located at $\omega-\omega_e =0$, i.e., $\omega=\omega_e$, as shown in Fig.~\ref{FigS3} (orange dashed line).
Substituting Eq.~\eqref{Jz_SCS} and Eq.~\eqref{Jz2_SCS} into Eq.~\eqref{Eq:omega uncertainty} and Eq.~\eqref{Eq:B uncertainty}, one can analytically obtain the measurement precisions $\Delta \omega$ and $\Delta B$.
They read
\begin{equation}\label{Delta_omega1}
\Delta{\omega}\!=\!\frac{\tilde{A}_{1}\pi}{2\gamma B}\!\frac{(\omega-\omega_{e})^2}{\left|\sin\left[(\omega-\omega_{e}) T\right]-T(\omega-\omega_{e})\cos\left[(\omega-\omega_{e}) T\right]\right|},
\end{equation}
\begin{equation}\label{Delta_B1}
\Delta{B}=\frac{\tilde{A}_{1}\pi}{2\gamma}\left|\frac{\omega-\omega_{e}}
{\sin\left[(\omega-\omega_{e}) T\right]}\right|.
\end{equation}
Here, the coefficient $\tilde{A}_{1}$ is
\begin{equation}\label{coefficientE}
\tilde{A}_{1}=\frac{\sqrt{\frac{1}{2}\sum_{m=-J}^{m=J}\!m^2|c_{m}^{J}(\theta)|^2\!-\!\left[\frac{1}{2}\!\sum_{m=-J}^{m=J}(-1)^{J\!-\!m}m|c_{m}^{J}(\theta)|^2 \textrm{sin}(2m\phi_1)\right]^2}}{|\sum_{m=-J}^{m=J}m^2(-1)^{J-m}|c_{m}^{J}(\theta)|^2 \cos(2m\phi_1)|}.
\end{equation}
Particularly, when $\phi_{1}=\pi/2$, $\sin(2m\phi_{1})=0$, and $(-1)^{J-m}\cos(2m\phi_{1})=1$, the measurement precisions $\Delta \omega$ and $\Delta B$ can be simplified as
\begin{eqnarray}\label{Delta_omega1}
\Delta{\omega}|_{\phi_{1}=\pi/2}&=&\frac{\pi}{2\gamma B}\frac{\sqrt{\frac{1}{2}\!\sum_{m=-J}^{m=J}\!m^2|c_{m}^{J}(\theta)|^2}}{|\sum_{m=-J}^{m=J}m^2|c_{m}^{J}(\theta)|^2|}\frac{(\omega-\omega_{e})^2}{\left|\sin\left[(\omega-\omega_{e}) T\right]-T(\omega-\omega_{e})\cos\left[(\omega-\omega_{e}) T\right]\right|}\nonumber \\
&=&{C(\theta)}\cdot\frac{\pi}{2\gamma B N}\frac{(\omega-\omega_{e})^2}{\left|\sin\left[(\omega-\omega_{e}) T\right]-T(\omega-\omega_{e})\cos\left[(\omega-\omega_{e}) T\right]\right|},
\end{eqnarray}
\begin{eqnarray}\label{Delta_B1}
\Delta{B}|_{\phi_{1}=\pi/2}&=&\frac{\sqrt{\frac{1}{2}\!\sum_{m=-J}^{m=J}\!m^2|c_{m}^{J}(\theta)|^2}}{|\sum_{m=-J}^{m=J}m^2|c_{m}^{J}(\theta)|^2|}\frac{\pi}{2\gamma}\left|\frac{\omega-\omega_{e}}
{\sin\left[(\omega-\omega_{e}) T\right]}\right| \nonumber \\
&=&C(\theta)\cdot\frac{\pi}{2\gamma{N}}\left|\frac{\omega-\omega_{e}}
{\sin\left[(\omega-\omega_{e}) T\right]}\right|.
\end{eqnarray}
Here, the coefficient $C(\theta)=1+\frac{2 \tan^{2}(\theta/2)}{1-\tan^{2}(\theta/2)}$ is dependent on particle number $N$.
It is indicated that, the measurement precisions $\Delta{\omega}$ and $\Delta B$ for spin cat states can exhibit the Heisenberg-limited scaling.
Further, we numerically calculate the optimal measurement precisions versus particle number in the range of $\left|(\omega-\omega_{e})/\omega\right|\in[0,0.008]$ and $\gamma B=2\pi$.
According to the fitting results (see Fig.~\ref{FigS5}), the log-log measurement precision for $\textrm{ln}\left({{\Delta \omega_{\textrm{min}}}/{\omega}}\right)\approx -\textrm{ln}({N})-8.26$ and $\textrm{ln}\left({\gamma \Delta B_{\textrm{min}}}/{\omega}\right)\approx -\textrm{ln}({N})-5.46$.
The optimal measurement precision $\Delta \omega_{\textrm{min}}$ (yellow square) and $\Delta B_{\textrm{min}}$ (yellow square) are both beyond SQL and approach to the Heisenberg-limited scaling.

At last, we also consider employing the GHZ state as the input state, i.e., $\theta=0$ in the spin cat state.
According to Eq.~\eqref{Evo_Spin_cat_state} and Eq.~\eqref{Jz_Spin_cat_state},  the final state $|\Psi\rangle_{\text{final}}^{I}$ is
%
%
\begin{eqnarray}\label{Evo_GHZ}
|\Psi\rangle_{\text{final}}^{I}
&=&\frac{1}{\sqrt{2}}(\textrm{cos}(J\phi_1)-\textrm{sin}(J\phi_1))\ket{J,J}\nonumber \\
&+&\frac{1}{\sqrt{2}}(\textrm{cos}(J\phi_1)+\textrm{sin}(J\phi_1))\ket{J,-J}.
\end{eqnarray}
The expectation of the half-population difference is
\begin{eqnarray}\label{Jz_GHZ}
\langle J_{z} \rangle_{\text{f}}=-\frac{N}{2}(-1)^L\textrm{sin}(N\phi_1).
\end{eqnarray}
Clearly, the main frequencies of the bi-sinusoidal oscillation of $\langle J_{z} \rangle_{\text{f}}$ becomes proportional to $N=2J$.
Similarly, we find the function of Eq.~\eqref{Jz_GHZ} with respect to detuning $\omega-\omega_{e}$ is symmetry under fixed $B$ and the symmetry point located at $\omega-\omega_{e}=0$, i.e., $\omega=\omega_e$, as shown in Fig.~\ref{FigS3}(blue lines).
Moreover, the square of half-population difference is independent on the two parameters, which becomes $\langle J_{z}^{2} \rangle_{\text{f}}={J^{2}}=N^2/4$.
According to Eq.~\eqref{Eq:omega uncertainty} and Eq.~\eqref{Eq:B uncertainty}, the measurement precisions for $\omega$ and $B$ has analytical form and they read,
\begin{equation}\label{GHZ_Delta_omega1}
\Delta{\omega}=\frac{\pi}{2N\gamma B}\frac{(\omega-\omega_{e})^2}{\left|\sin\left[(\omega-\omega_{e}) T\right]-T(\omega-\omega_{e})\cos\left[(\omega-\omega_{e}) T\right]\right|},
\end{equation}

\begin{equation}\label{GHZ_Delta_B1}
\Delta{B}=\frac{\pi}{2N\gamma}\left|\frac{\omega-\omega_{e}}
{\sin\left[(\omega-\omega_{e}) T\right]}\right|.
\end{equation}
From Eq.~\eqref{GHZ_Delta_omega1} and Eq.~\eqref{GHZ_Delta_B1}, the measurement precisions $\Delta{\omega}$ and $\Delta B$ can exhibit Heisenberg-limited scaling.
Meanwhile, for a fixed $N$, the measurement precision $\Delta \omega$ is dependent on both detuning $\omega-\omega_{e}$ and $B$, as show in Fig.~\ref{FigS4}~(c).
When $\sin[(\omega-\omega_{e})T](2-(\omega-\omega_{e})^2T^{2})-2T(\omega-\omega_{e})\cos[(\omega-\omega_{e})T]=0$, $\Delta \omega$ attains its optimal value.
However, the measurement precision $\Delta B$ is just dependent on detuning $\omega-\omega_{e}$.
When it is near resonance $\omega\approx\omega_{e}$, $\Delta{B}=\frac{\pi}{2\gamma\sqrt{N}T}$ is optimal.
To confirm the dependence of measurement precisions on particle number, we numerically calculate the optimal measurement precisions verse particle number in the range of $|(\omega-\omega_{e})/\omega|\in[0,0.008]$ and $\gamma B=2\pi$.
According to the fitting results (see Fig.~\ref{FigS5}), the log-log measurement precision for $\textrm{ln}\left({{\Delta \omega_{\textrm{min}}}/{\omega}}\right)\approx -\textrm{ln}({N})-8.37$ and $\textrm{ln}\left({\gamma \Delta B_{\textrm{min}}}/{\omega}\right)\approx -\textrm{ln}({N})-5.53$.
For the two parameters, the optimal measurement precision $\Delta \omega_{\textrm{min}}$ (pink triangle) and $\Delta B_{\textrm{min}}$ (pink triangle) both exhibit the Heisenberg-limited scaling.
\end{widetext}

\end{document}